# Bi-stable Nonlinear Energy Sinks (BNESs) for Response Mitigation and Drag Reduction of Subsea Cables Undergoing Vortex-induced Vibrations

Anargyros Michaloliakos[a*], Ryan Davies [b], Matthew Hall [c], Alexander F. Vakakis[a]

[a] *University of Illinois, Urbana, Urbana, Illinois, USA*
[b] *Stanford University, Stanford, California, USA*
[c] *SINTEF Ocean, Trondheim, Norway*
**Corresponding author, anar.michaloliakos@gmail.com*

## Abstract

A methodology for passive vortex-induced vibration (VIV) mitigation of subsea dynamic power cables by means of a set of optimized strongly nonlinear bi-stable mass-spring-damper attachments, referred as bi-stable nonlinear energy sinks (BNESs) is studied within the open-source MoorDyn library. A fully 3D computational framework is formulated, capturing realistic cable dynamics, hydrodynamic forcing via Morison-type models, and nonlinear passive vibration-mitigation mechanisms under spatially and temporally varying dynamic currents. The BNESs are integrated into this time-domain cable dynamics computational model in a consistent way that naturally accommodates highly non-stationary VIVs. An extensive data-driven optimization study is conducted by systematically sampling the BNES design space across multiple dynamic current profiles to demonstrate that properly tuned BNES configurations can achieve substantial reductions in peak-to-peak cable VIV amplitudes. Perhaps equally important, the actions of the BNESs lead to significant and robust decrease of the cumulative VIV-induced energy intake to the cable by the surrounding flow, as well as analogous reduction of the drag energy experienced by the cable due to VIV. To the authors' best knowledge, the capacity to reduce energy intake and drag amplification in a subsea cable using passive nonlinear attachments is reported for the first time in the literature and can be particularly beneficial for reducing short- and long-term fatigue in these systems. Time–frequency wavelet postprocessing analysis clarifies the physical nonlinear mechanisms underlying the reported effective VIV mitigation by the optimized BNESs, and reveals a consistent pattern of targeted nonlinear energy transfers and VIV energy scattering from high-amplitude, low-frequency cable modes (which are dominant in the baseline cable without attached BNESs undergoing VIV), to lower-amplitude, higher frequency cable modes. This energy scattering (modal redistribution) facilitates rapid dissipation through hydrodynamic damping and internal structural losses of the cable vibrations, explaining the simultaneous reduction in vibration amplitudes and the cumulative energy intake in the cable from the flow. This study demonstrates the efficacy of BNESs for effective and robust mitigation of VIV of subsea power cables under realistic unsteady operating conditions and paves the way for future investigations of combined current–wave loading, platform-induced motions, fatigue life estimation and betterment through VIV reduction in this and other types of marine structures.

## 1. Introduction

Subsea power cables are critical components of floating offshore wind farms and almost always operate in harsh underwater environments where highly nonstationary loads can produce vibration and fatigue. Dynamic subsea cables extend freely through the water column and are particularly vulnerable, as they are subject to continuous platform motion and excitations such as tidal currents and waves, which induce excess vibrations leading to fatigue, necessitating costly repairs and undermining the economic viability of offshore wind farms. Therefore, addressing the stable operation, reliability and durability of dynamic subsea cables is crucial for enhancing the resilience and efficiency of floating offshore wind energy transmission systems, and so there is a need for efficient and robust methods for addressing such unwanted effects to ensure extension of service life with feasible maintenance costs.

In that context, vortex-induced vibrations (VIVs) of bluff bodies is a well-studied fluid-structure phenomenon caused when steady or unsteady sea currents interact with these subsea systems (Williamson and Govardhan, 2004; Paidoussis et al., 2011). Regarding methods specifically for simulating VIVs in subsea cables, given that typically these are highly non-stationary unsteady fluid-induced structural oscillations, there is a need of their study in the time domain. To this end, there are limited time-domain models and methods for investigating VIVs in practical subsea cables (Thorsen, 2016; Wu et al., 2020; Anderson et al., 2024, Davies et al., 2026). These are based on a semi-empirical framework first presented in Thorsen (2016), with the most prominent implementation being the recently released commercial model VIVANA-TD (Wu et al., 2020). There also exist commercial frequency-domain models for simulating VIV in cables (e.g., VIVANA, SHEAR7, VIVA), but these cannot capture complex, unsteady and non-stationary dynamics in the time-domain (Boru, 2021). While there has been work modeling VIV through high-fidelity computational fluid dynamics, the required computational cost is prohibitive for application to iterative analysis of subsea power cables, e.g., (Bourguet et al., 2011; Tumkur et al., 2017; Blanchard et al., 2020, Jiang et al. 2024). In the other limit of low fidelity modeling, low-dimensional reduced-order models such as wake oscillators or other finite-order models have been developed to study fluid-structure interaction and VIV, but their applicability is rather limited to simple geometries, relatively simple bluff bodies and restricted operating ranges, e.g., (Facchinetti et al., 2004; Tumkur et al., 2013).

Yet, accurately predicting VIV in practical subsea cables is of critical engineering importance. This is necessary not only for addressing the possibility for high-amplitude unstable cable vibrations, but also due to the added lift and drag forces that are generated over the entire length of the cable, which need to be accurately accounted for stable operation and safe maintenance. This holds especially when the vortex shedding frequency of the fluid becomes close to one of the natural frequencies of the modes of the cable, in which case sustained or transient resonance (or “lock-in”) may occur in a synchronization that causes large-amplitude oscillations (Paidoussis, 2014). There have been several recent analyses of dynamic power cable VIV. Elrick and Venugopal (2023) simulated VIV of a dynamic power cable attached to a floating wind turbine using commercial software and showed sensitivity of cable vibrations to system parameters and dynamic current profiles. Moideen et al. (2024a,b) completed a scaled tank testing campaign of a lazy-wave dynamic power cable in uniform currents and found evidence of large drag amplification. Davies et al. (2026) simulated at uni-directional wind, wave, and current excitation of a floating wind turbine and found that cable VIV can violate design factors of safety in extreme conditions.

Recently, Davies et al. (2026) successfully integrated cross-flow VIV simulation capacity into the broader dynamics computational model MoorDyn (Hall and Goupee, 2015), thus enabling capability for fully capturing cross-flow VIV of a dynamic power cable subject to realistic uni-directional unsteady dynamic currents. MoorDyn is open-source and can be coupled to other dynamical computational codes, thus constituting a powerful open-source numerical tool for accurately simulating VIV of subsea power cable dynamics. This capacity was validated in (Davies et al., 2026) against previous numerical models (Thorsen et al., 2017) and experimental results (Moideen et al. 2024). In the present work MoorDyn will be used as part of an optimization numerical framework for robustly and effectively suppressing VIV of subsea power cables by means of a set of strongly nonlinear local attachments. However, before discussing this nonlinear mitigation framework, a brief review of current approaches to mitigate subsea cable vibrations is given.

There have been developed different ways for passively mitigating the vibration of subsea structures, e.g., risers and cables, based on the structural augmentation by external lightweight local attachments, including helical strakes, streamlined fairings, and ribbons (Zheng and Wang, 2017; Drumond et al., 2018; Assi et al., 2022). In addition, local resonating attachments in the form of linear vibration absorbers have been considered, commonly referred to as tuned mass dampers (TMDs), e.g., (Ormondroyd and Den Hartog, 1928; Gutierrez Soto and Adeli, 2013; Elias and Matsagar, 2017). Although TMDs are effective in suppressing narrowband cable vibrations involving single-mode excitations, they are not as effective for multi-modal, multi-frequency vibrations, which are typical in unsteady VIV events, where the cable responses are highly non-stationary. In addition, even in cases where TMDs are effective in suppressing narrowband cable vibrations, their resulting oscillations are large-amplitude, which imposes impractical design constraints, excessive stresses leading to fatigue and demanding maintenance requirements. Lastly, an inevitable challenge associated with subsea cable operation is biofouling (Marty et al., 2021), which typically detunes TMDs from their optimal operating frequency range, significantly deteriorating their vibration mitigation capacity over the cable's service life.

To alleviate these limitations of the TMDs, an alternative class of nonlinear passive attachments have been considered, referred to as *nonlinear energy sinks (NESs)*, with capacity for broadband, multi-frequency vibration mitigation. These are passive, lightweight, dissipative local attachments with inherent strong nonlinearity, which, when attached to a primary structure, can absorb vibration energy in a broadband fashion (Vakakis et al., 2008). This is since, unlike TMDs, the NESs do not have preset resonance frequencies, so their mitigation effectiveness is based on the passive adaptation of their dynamics to the frequency and energy content of the applied excitations (Starosvetsky and Gendelman, 2010; Vakakis et al., 2022). This renders NESs particularly suited for mitigating highly non-stationary vibrations across extensive frequency spectra, which are typical in VIVs of subsea cables due to unsteady dynamic currents. This was demonstrated in (Michaloliakos et al., 2025; Michaloliakos, 2026) where bi-stable NESs (BNESs) were optimized for mitigating platform-motion-induced vibrations of subsea cables in still fluid, without accounting for vortex-induced dynamical excitations. The bi-stable nature of these attachments allows them to switch between two stable equilibrium states, facilitating a range of resonant interactions that enhance energy absorption from a primary structure (in this case the subsea dynamic power cable) and local energy dissipation (Manevitch et al., 2014; Romeo et al., 2015).

Various essentially nonlinear stiffness characteristics have been employed to realize NESs, including purely cubic (hardening) springs (Gendelman et al., 2001; Vakakis et al., 2008), combined softening–hardening geometric nonlinearities (Mojahed et al., 2018), bi-stable configurations with negative linear stiffness (Al-Shudeifat, 2014; Manevitch et al., 2014; Romeo et al., 2015), tri-stable configurations (Zeng and Ding, 2023), and non-smooth vibro-impact attachments (Nucera et al., 2007; Gourc et al., 2015). Among these, the bi-stable design is attractive for the strongly non-stationary excitations considered herein, since cross-well transitions can be activated at relatively low energies, providing a rich set of resonant interactions and strong energy scattering. Bi-stability, however, may also give rise to chaotic responses (Romeo et al., 2015), whose sensitivity to initial conditions may be undesirable in certain regimes; this aspect is discussed further in Section 4.3.

The scope of the present work is to demonstrate the efficacy of the BNES for robust mitigation of highly nonstationary VIVs in subsea dynamic power cables. Through a data-driven multi-objective optimization framework, a set of BNESs attached to a cable undergoing VIV is designed to maximize VIV suppression by rapidly absorbing, dissipating, and nonlinearly scattering vibration energy across the cable. This is performed by leveraging a validated time-domain VIV model directly into the open-source MoorDyn library (Davies et al., 2026), which enables dynamic current- and motion-induced VIV simulations in a fully open-source environment. This unified framework provides high-resolution transient responses, captures multi-harmonic cable excitations, and provides direct access to fatigue-relevant metrics—while also allowing seamless integration and evaluation of the mitigation effectiveness of the attached BNESs. Overall, this work presents a coupled modeling and optimization approach that both advances dynamic cable simulation capabilities and demonstrates the strong potential of BNES devices as passive solutions for enhancing the durability and performance of subsea power cables.

## 2. VIV Modeling in MoorDyn

In this work, the lumped-mass mooring dynamics model MoorDyn is applied for the simulation of a subsea dynamic power cable with an attached set of BNESs undergoing VIV. MoorDyn is an open-source time-domain mooring dynamics library integrated into numerous floating renewable energy simulators, including the floating wind turbine simulation tool OpenFAST (Hall and Goupee, 2015). MoorDyn has the capability to model rigid-body dynamics, shared mooring lines, line failures, and dynamic power cables, having four core objects that can be used to construct a mooring system: Lines, points, rods, and bodies. MoorDyn employs lumped-mass discretization to represent mooring lines as point masses connected by linear spring-damper segments, which allows for elasticity in the axial direction. It can also model bending stiffness based on the relative alignment of successive line segments and transmits bending moments between different bodies using rod objects (Hall, 2020). Figure 1 shows elements of the model, where $\boldsymbol{r_i}$ represents the absolute position vector of node $i$ along the cable, as first applied by Michaloliakos et al. (2025) for NESs on dynamic cables.

A recent implementation of transient vortex-induced vibration capabilities within the open-source MoorDyn computational framework has been reported by Davies et al. (2026). This implementation is based on the time-domain cross-flow VIV model originally developed by (Thorsen, 2016), with adaptations to ensure compatibility with MoorDyn's lumped-mass formulation for flexible marine structures. Thorsen's model provides a time-domain representation of cross-flow VIV forces that, when coupled with a lumped-mass cable model, accurately captures the dynamic response of submerged cylindrical structures subjected to both steady and spatially or temporally varying dynamic currents (Thorsen et al., 2026, 2017). In this formulation, the cross-flow VIV force acting on the cable is modeled as a point force applied at each hydrodynamic node. The force depends on the local transverse relative fluid velocity and incorporates a synchronization model to represent the coupling between structural motion and vortex shedding. The VIV force is superimposed on the baseline hydrodynamic loading, which in MoorDyn is computed using Morison's equation, as described in (Thorsen et al., 2017; Michaloliakos, 2026). Since MoorDyn already accounts for inertia and drag forces through Morison-type formulations for line objects (Michaloliakos et al., 2025; Michaloliakos, 2026), only the lift component associated with cross flow VIV needs to be described briefly, and more details can be found in (Davies et al., 2026).

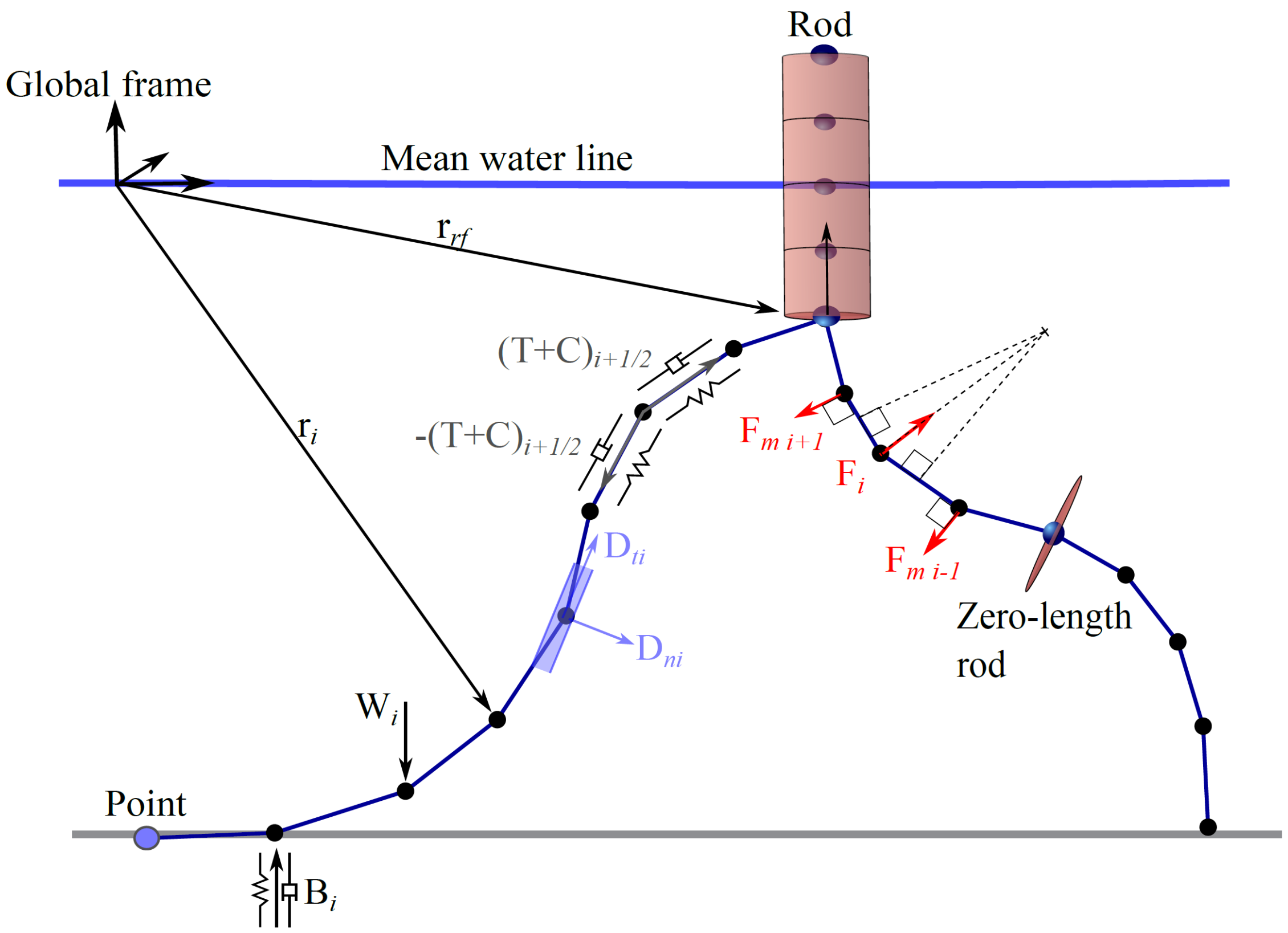

Figure 1: Schematic of the MoorDyn lumped mass model of subsea dynamic power cable (Michaloliakos et al., 2025).

It is noted that, although each segment of the lumped-mass discretization obeys a linear axial elastic law with linear internal damping, the cable model as a whole is geometrically nonlinear: The nodal positions are tracked in three dimensions, the segment tensions and directions are recomputed at every time step from the current deformed configuration, and bending stiffness enters through the relative rotation of adjacent segments (Hall and Goupee, 2015; Hall, 2020). Hence, large displacements and rotations, tension–geometry coupling, seabed contact and the coupling between axial, transverse and bending responses are captured, whereas material nonlinearity is not considered, which is consistent with power cables operating in their elastic range. MoorDyn has been validated against experimental mooring-line data (Hall and Goupee, 2015), the implementation of the BNES force–deflection law has been validated against the analytical model (Michaloliakos et al., 2025), and the VIV extension has been validated against the numerical results of Thorsen et al. (2017) and the tank tests of Moideen et al. (2024) in Davies et al. (2026).

Accordingly, the cross-flow lift vector force acting on node $i$ of the cable, denoted by $\boldsymbol{F}_{VIV}$, is expressed as,

$$\boldsymbol{F}_{VIV} = \left(\frac{1}{2}\right) \rho_w \, d \, \left|\boldsymbol{v}_p\right| C_l \, cos\phi_l \left(\hat{\boldsymbol{q}} \times \boldsymbol{v}_p\right) \tag{1}$$

where $\rho_w$ is the fluid density, $d$ the cable diameter, $C_l$ the non-dimensional lift coefficient, $\phi_l$ the phase of the lift force, and $\boldsymbol{v}_p$ the vector of transverse relative velocity of the cable at node $i$. Here, $\hat{\boldsymbol{q}}$ denotes the unit vector tangent to the cable at node $i$, so that the cross product $\hat{\boldsymbol{q}} \times \boldsymbol{v}_p$ is normal to both the cable axis and the local relative flow and defines the cross-flow direction of the lift force. This geometric definition ensures that the lift force direction is correctly captured for arbitrary cable orientations and flow directions, enabling the simulation of complex loading scenarios, including heave-induced vortex-induced vibrations. The magnitude of the lift force scales with the square of the transverse relative velocity, indicating that energy is transferred from the surrounding fluid into the cable through the VIV excitation mechanism. The lift coefficient $C_l$ depends on the Reynolds number of the given application, and this is associated with the local flow conditions; hence, it must be selected consistently with the intended excitation regime, particularly when calibrating the model against experimental measurements or field data. A more detailed discussion regarding the selection of the lift coefficient can be found in (Davies et al., 2026; Michaloliakos, 2026), and Thorsen's model is implemented in the present VIV cable study by selecting the lift coefficient such that the vibration amplitudes predicted by the MoorDyn simulations are consistent with experimental results for comparable systems. Hence, a lift coefficient $C_l = 1.2$ was used herein, and future improvements to the VIV model could incorporate a lift coefficient that varies with reduced velocity, thereby enhancing predictive capability under unsteady flow conditions.

Two remarks regarding Eq. (1) are in order. First, the depth- and time-dependence of the VIV forcing enters through the local relative velocity $\boldsymbol{v}_p$, which varies along the span according to the imposed current profile; the force magnitude scales with $|\boldsymbol{v}_p|^2$, and the local vortex-shedding (synchronization) frequency also varies along the cable, whereas $C_l$ is a model constant that sets the amplitude of the lift force. Second, the fluid–structure coupling is two-way at the level of the model: The lift force depends on the instantaneous relative velocity between fluid and cable, and its phase $\phi_l$ is continuously synchronized to the structural motion, so that the cable response feeds back on the hydrodynamic force at every time step and lock-in emerges dynamically; similarly, the Morison drag and added-mass terms are evaluated with the relative fluid–structure kinematics. What the semi-empirical model does not resolve is the wake flow field itself, which would require computational fluid dynamics at a cost that is prohibitive for the tens of thousands of simulations of the optimization study that follows. Regarding the value of the lift coefficient, values ranging from $C_l$ = 0.65 (Ulveseter et al., 2018) to the nominal bare-cylinder value $C_l$ = 1.2 (Wu et al., 2020) have been adopted in the literature, while Bayesian optimization (Andersen et al., 2024) and CFD-based tuning (Jiang et al., 2024) have also been proposed for its calibration; the value $C_l$ = 1.2 adopted herein follows Wu et al. (2020) (see also Michaloliakos, 2026).

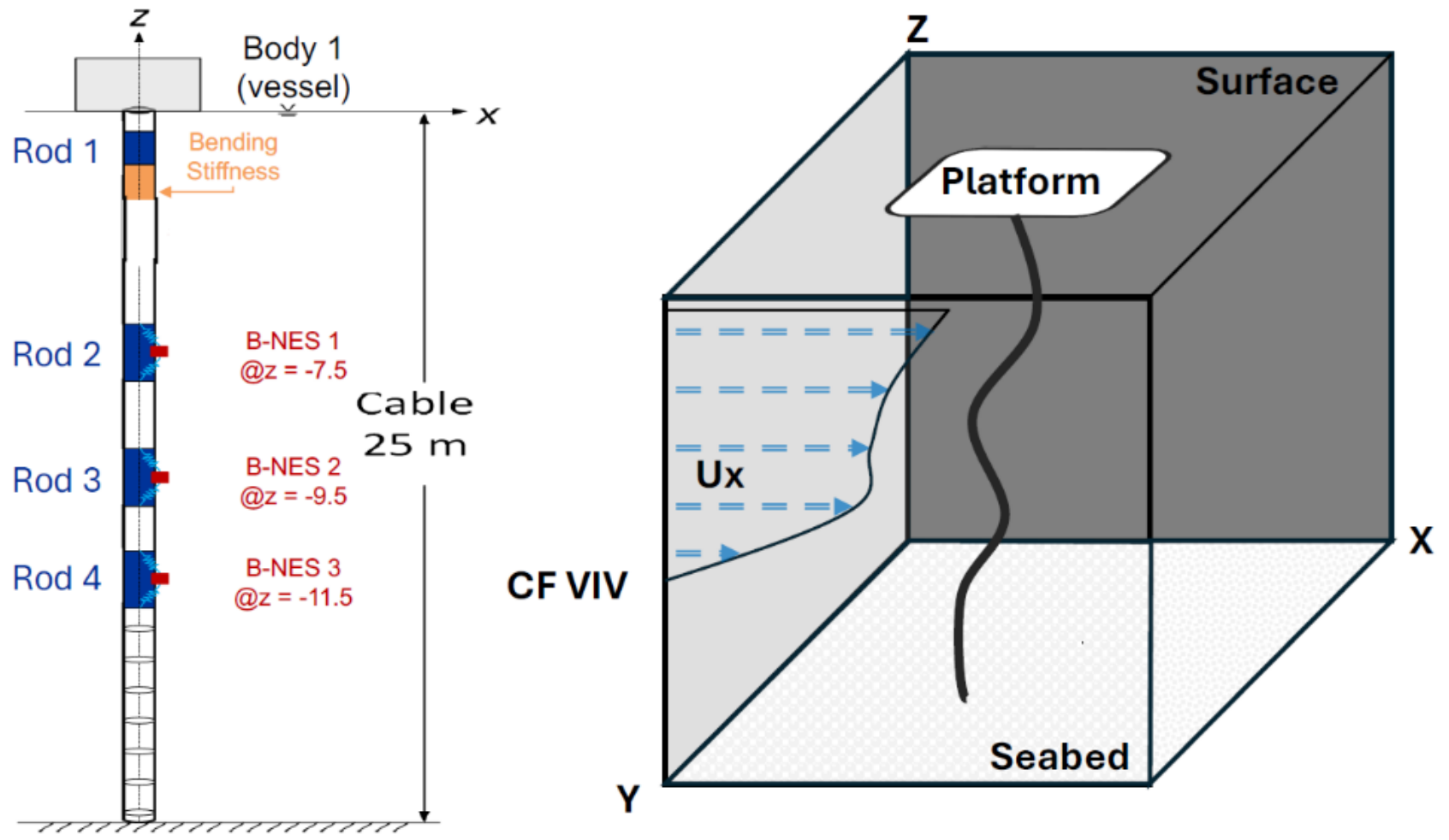

(a) (b)

Figure 2. Schematics of (a) the initially (at $t = 0$) vertical subsea power cable extending from a floating platform to the seabed, with three attached BNESs (labeled from 1 to 3) at prescribed depths along its span; (b) imposed non-uniform, time-varying dynamic current acting in the $x$ −direction, and cross-flow direction in the $(y, z)$ plane of vortex-induced vibrations; the resulting cable motion occurs primarily in the plane perpendicular to the current direction.

## 3. System Modeling in MoorDyn

Following (Michaloliakos et al., 2025), we adopt the same subsea cable configuration and BNES attachment strategy to investigate the VIV cable mitigation performance. The cable has a diameter of 0.076 m and a mass density of 4.73 $Kg/m$. The axial stiffness ($EA$) of the cable is set equal to 9 $MN$, and its bending stiffness ($EI$) to 71 $kN\ m^2$. The added-mass coefficient normal to the cable, describing the influence of the surrounding fluid on the cable's inertia, is set to Can = 1.0. The normal and tangential hydrodynamic drag coefficients are set to Cdn = 1.2 and Cdt = 0.0, respectively (Michaloliakos et al., 2025). The subsea cable is modeled in MoorDyn with as initial vertical line (at $t = 0$) extending from a floating platform to the seabed. All geometric, structural, and hydrodynamic parameters of the cable are identical to those in (Michaloliakos et al., 2025).

In contrast to that previous study which assumed still-water and platform-driven excitation, the present study considers a spatially and temporally varying dynamic current, which induces cross-flow VIV along the cable span. This is modeled according to the numerical scheme discussed in the previous section. The dynamic current is prescribed in the global $x$ −direction and varies with both depth and time, generating lift forces that excite transverse oscillations of the cable predominantly in the $(y, z)$ plane, i.e., perpendicular to the flow direction. This excitation mechanism results in fully three-dimensional (3D) cable motion, with coupled axial, transverse, and bending responses. Because VIV dynamics are largely dominated by cross-flow motion, a cross-flow-only model was sufficient for the analysis of the BNESs (Thorsen 2016, Moideen et al. 2024a). Figure 2a illustrates the modeling configuration employed in this study, with the initially vertical subsea power cable attached to the platform, along with the locations of three BNESs distributed along its span. Figure 2b provides a schematic representation of the applied dynamic current field and the corresponding definition of the cross-flow direction used to characterize VIV-induced cable motion.

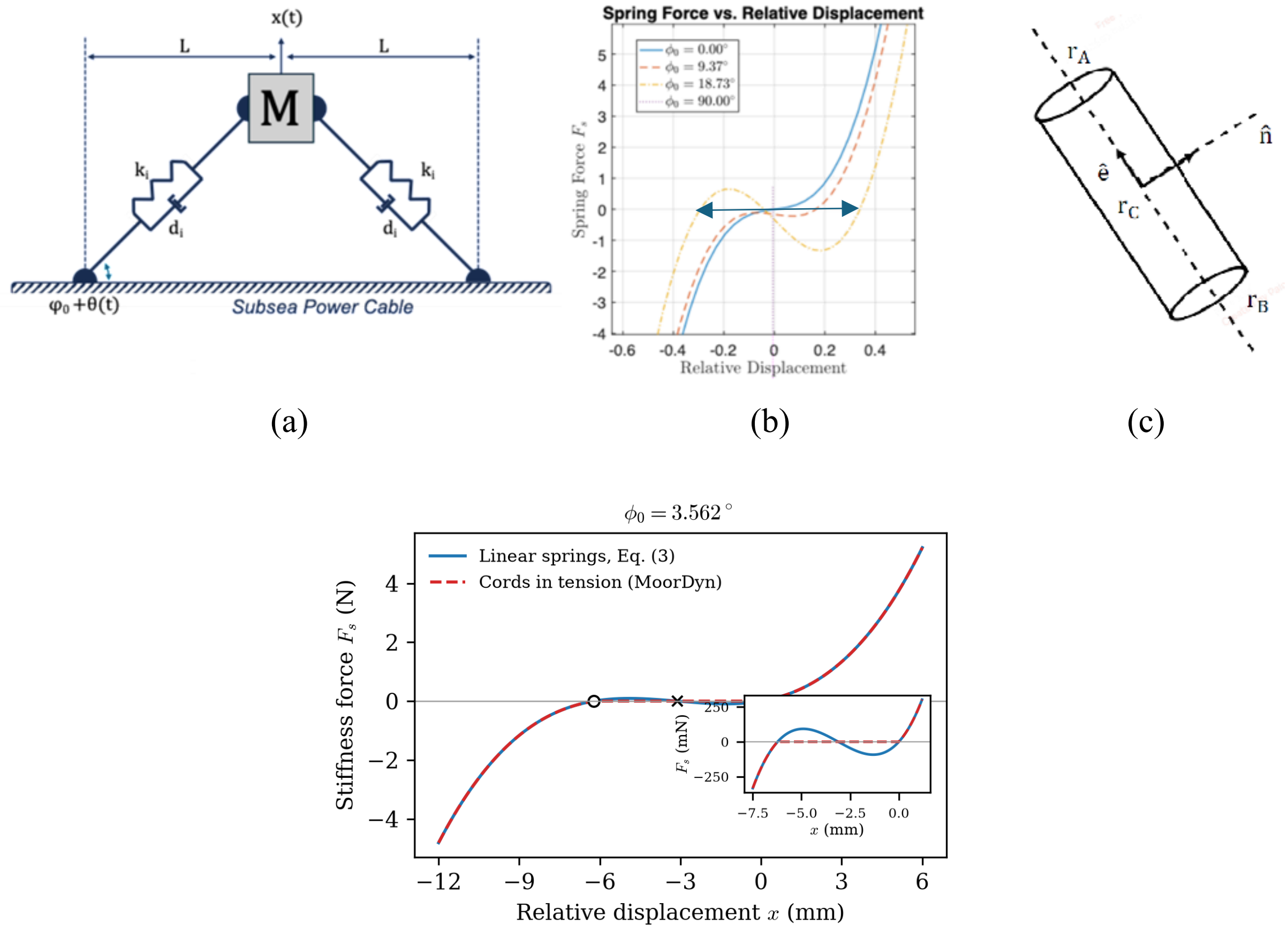

(d)

Figure 3. Bi-stable nonlinear energy sink (BNES) attached to the subsea cable: (a) System configuration with $\phi_0$ being the initial angle of inclination at equilibrium – the pair of inclined elements are linearly elastic chords-in-tension; (b) tunability of the nonlinear stiffness force exerted by the BNES with respect to $\phi_0$ (bi-stability "jumps" are shown for $\phi_0 = 18.73°$); (c) the rod element on the cable acting as basis for attaching the BNES in the MoorDyn implementation, enabling a SDOF model for the BNES while preserving fully 3D cable motion ($\mathbf{r}_A$, $\mathbf{r}_B$ and $\mathbf{r}_C$ are absolute positions of the ends and center of the rod, respectively, while $\hat{\boldsymbol{e}}$ and $\hat{\boldsymbol{n}}$ are the local unit vectors in directions along and transversely to the rod, respectively); and (d) stiffness force law of the BNES at the operating angle $\phi_0 = 3.562°$ adopted in the following optimization study, for the linear-spring model (3) and the cords-in-tension implemented in MoorDyn, with the parameters of (b) – open circles denote the stable equilibria, the cross the unstable one, and the inset magnifies the bi-stable region.

This setup represents, to the authors' best knowledge, the first fully integrated numerical framework combining 3D subsea cable dynamics, time-varying depth dependent VIV excitation, and attached nonlinear passive vibration-mitigation devices. The formulation is general and readily extendable to a wide range of environmental conditions, cable geometries, and nonlinear attachment designs. The main modelling assumptions and simplifications of the present study are summarized as follows: (i) Only cross-flow VIV is modelled, and in-line VIV being neglected; (ii) the VIV force follows the semi-empirical model of Thorsen (2016), with a constant lift coefficient $C_l = 1.2$ and the synchronization parameters of Davies et al. (2026); (iii) inertia and drag forces are computed with Morison-type formulations based on the relative fluid–structure kinematics; (iv) the cable is linearly elastic, with full geometric nonlinearity; (v) the cable is initially vertical, fixed at the seabed and attached to a platform that is held fixed, i.e., wave loading and platform motions are not considered; (vi) each BNES is modelled as a single-degree-of-freedom mass constrained to move along a rod-local transverse direction and connected to the cable through linearly elastic cords that carry tension only, with linear axial damping; and (vii) the total mass of the BNES-equipped cable is matched to that of the baseline cable.

A brief description of the attached BNESs is now given, and more details can be found in (Michaloliakos et al., 2025). Each BNES is modeled as a nonlinear single-degree-of-freedom (SDOF) system with a mass connected to the subsea cable through two inclined pairs of elastic-cords-in-tension with nominal stiffness $k_i$ and damping coefficient $d_i$ (Fig. 3a). At equilibrium the inclined elements have an initial angle of inclination $\phi_0$ relative to the cable, and even though the inclined chord elements are linearly elastic, strong geometric nonlinearity is generated with the relative motion of the mass of the BNES with respect to the cable (Mojahed et al., 2018). The angle of inclination significantly influences its nonlinear response. Indeed, assuming that the mass $M$ of the BNES oscillates orthogonally and with relative displacement $x$ with respect to the cable, its governing equation of motion is given by:

$$M(\ddot{x} + \ddot{x}_{cable}) + F_{si}(x) + F_{di}(x, \dot{x}) = 0 \qquad (2)$$

where $x_{cable}$ is the local motion of the cable at the points of attachment to the BNES, and $F_{si}(x), F_{di}(x, \dot{x})$ are the generated nonlinear stiffness and dissipative forces (Mojahed et al., 2018),

$$F_s = 2k_i\left(1 - \frac{L\sec(\phi_0)}{\sqrt{L^2 + y^2}}\right)y, \qquad F_{di} = 2d_i\frac{\dot{y}y^2}{L^2 + y^2} \qquad (3)$$

where $y = x + L\tan\phi_0$. Depending on the initial angle of inclination at equilibrium $\phi_0$, different nonlinear stiffness laws are achieved (Fig. 3b), for system parameters, $M = 0.1\,Kg, k_i = 2 \times 10^4\,N/m\,, d_i = 0.25\,Ns/m\,, L = 0.05m$. Indeed, above a critical threshold of $\phi_0$ *bi-stability* occurs with two non-trivial stable equilibrium positions and an unstable trivial one (Mojahed et al., 2018). For the BNES geometry considered herein the bifurcation value is $\phi_{0,bif} = 0.22°$ (Michaloliakos et al., 2025; Michaloliakos, 2026), so the operating angle $\phi_0 = 3.562°$ adopted in Section 4 lies well inside the bi-stable regime, as shown by the stiffness law of Fig. 3d; sudden transitions ("jumps") of the BNES mass between the two stable equilibria generate strong nonlinearity which drastically enhance the vibration mitigation performance of the BNES and increases its capacity to absorb, scatter and dissipate a significant part of the VIV-induced energy of the cable. At this point, two points are emphasized: (i) Since the inclined chords can only exert force while in tension (and lose this ability when they are compressed), when the BNES mass passes through static equilibrium there is an instantaneous bi-stability transition – this is shown in Fig. 3b; and (ii) to reduce computational complexity in the numerical optimization study that follows, the dissipative force in (3) is realized in MoorDyn by assigning to each inclined cord a linear viscous (axial) damping coefficient that is proportional to its axial stiffness. Since the elongation rate of each cord is $y\dot{y}/(L^2 + y^2)^{1/2}$, the projection of the resulting cord forces onto the direction of motion of the BNES mass is proportional to $y^2\dot{y}/(L^2 + y^2)$, i.e., the geometric nonlinear damping of (3) is retained in the simulations. For the optimized designs of Section 4 the half-length of the attachment rod ($L \approx 0.5$ m) greatly exceeds the BNES strokes (at most 2–5 cm), so that the damping operates in the geometrically nonlinear regime $y \ll L$, which has been shown to enhance the robustness of nonlinear absorbers (Zhang et al., 2021; Dekemele, 2023).

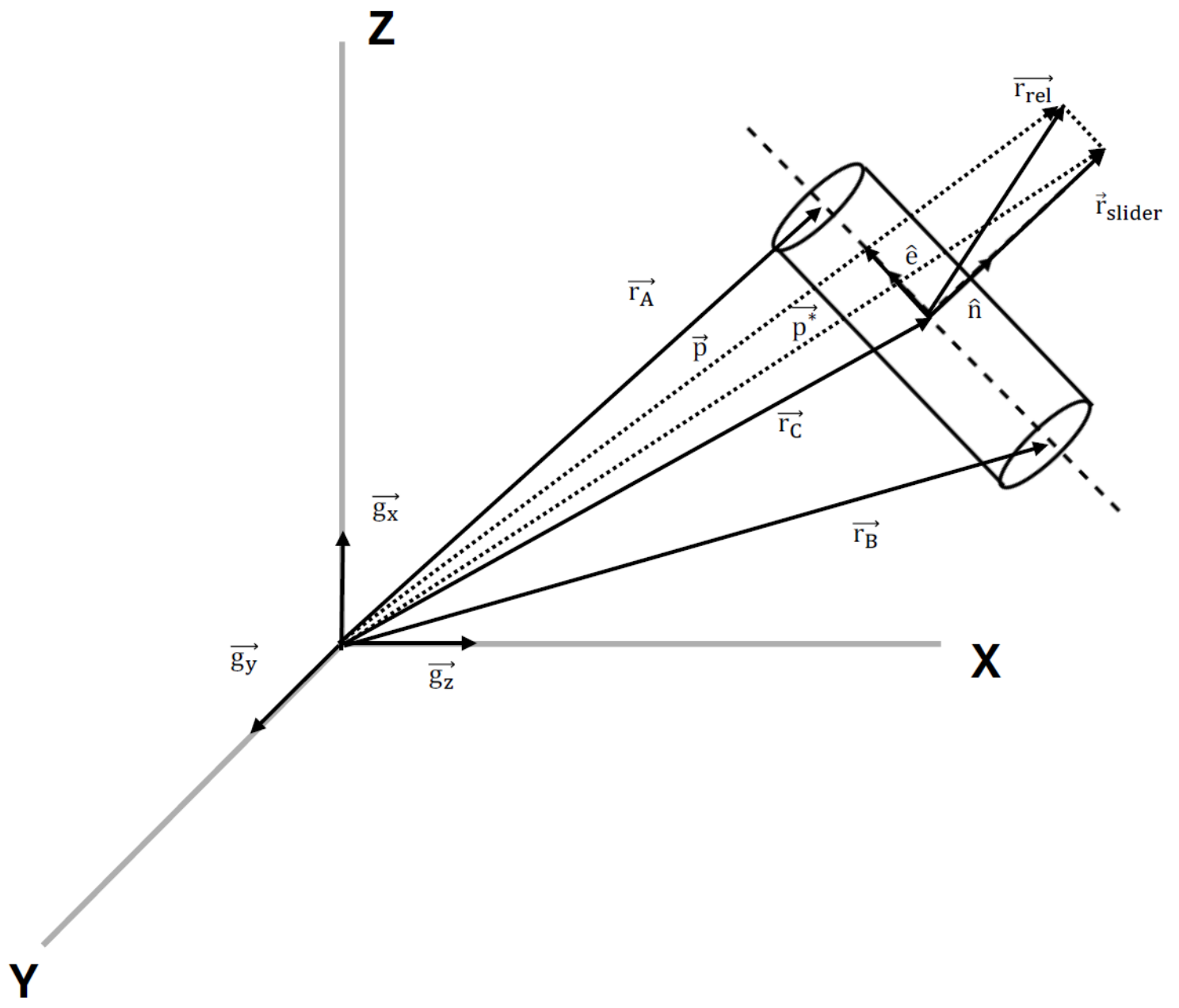

Figure 4. Geometry of the MoorDyn numerical implementation of the BNES and the rod constraint at its base (along the cable), showing the rod end nodes ($A$, $B$), center ($C$), and local unit vectors in directions along ($\hat{\boldsymbol{e}}$) and transverse ($\hat{\boldsymbol{n}}$) to the rod axis; the unconstrained and constrained BNES mass positions $\boldsymbol{p}$ and $\boldsymbol{p}^*$, and corresponding BNES mass velocities are shown in the global reference frame.

The bi-stability effect is manipulated by varying the ratio of the natural lengths of the springs relative to the rod length: The smaller this ratio is, the greater the angle of inclination becomes, thus enhancing bi-stability. The unique aspect of this setup is the clearance effect introduced by the linear inclined elastic cords in-tension, which do not exert force unless the displacement of the BNES mass relative to its attachment point exceeds their natural lengths (see Fig. 3b). This feature introduces a significant non-smooth nonlinearity that allows for the exploitation of rich dynamics. It is emphasized that the cords are not pre-tensioned: At the equilibrium position they are at (or marginally below) their natural length and exert no force. Because they act only in tension, the barrier region of the bi-stable potential of the linear-spring model, located between its two stable equilibria, is replaced by a clearance band of width $2L \tan\phi_0$ within which the cords are slack (Fig. 3d); the transitions of the BNES mass across this band constitute the cross-well transitions discussed in Section 4.3.

Regarding the numerical modeling of the attached BNESs to the cable, following (Michaloliakos et al., 2025; Michaloliakos, 2026), a rod element is added to the cable, extending over the length defined by the two points of attachment of the BNES at absolute positions $\mathbf{r}_A$ and $\mathbf{r}_B$, as illustrated in Figs. 3c and 4. A key extension introduced in this work concerns the kinematic treatment of the BNES mass in the presence of fully 3D cable motion. In earlier planar formulations, in the absence of VIV the relative motion of the BNES was implicitly confined to be 1D on a fixed plane (Michaloliakos et al., 2025). However, under VIV excitation, the subsea cable undergoes general 3D motion, rendering such simplifying assumptions invalid. To ensure realistic BNES motion while accommodating the more complex cable dynamics, the BNES mass is modeled as a MoorDyn point object that is kinematically constrained to move along a single rod-local transverse direction, while the supporting rod itself is free to translate and rotate in three dimensions together with the cable (see Fig. 2b). All geometric quantities defining this three-dimensional constraint are illustrated schematically in Fig. 4. This formulation enforces a single geometric degree of freedom for the BNES mass – consistent with the previous analytical SDOF model – while fully preserving 3D coupling between the cable and the nonlinear attachment. By exploiting the open-source and extensible nature of the MoorDyn framework, this approach enables the systematic investigation and optimization of nonlinear passive vibration-mitigation of cable VIV subject to realistic subsea conditions characterized by complex 3D fluid-structure interaction dynamics. For more details of the numerical implementation of the BNESs to MoorDyn one is referred to the Thesis by Michaloliakos (2026).

## 4. Numerical Optimization of the set of BNESs for VIV Mitigation

### 4.1. Numerical Optimization Framework

An advanced data-driven optimization framework is discussed in this section to maximize the VIV mitigation performance of the set of BNESs that are attached to the subsea power cable, building upon previous work (Michaloliakos et al., 2025). The objective of this study is to systematically identify the BNES configurations and system parameters that maximize suppression of VIV of the cable, under realistic excitation scenarios, while accounting for the strongly nonlinear and transient nature of the coupled cable–BNES dynamics. Rather than relying on large-scale high-performance computing infrastructure, the present study employs parallel computing techniques to efficiently explore the high-dimensional design space associated with the BNES parameters. To this end, independent time-domain simulations are distributed across multiple CPU cores, enabling simultaneous evaluation of candidate designs. This strategy substantially reduces wall-clock time while maintaining full numerical fidelity and is particularly effective for handling the (numerically) stiff and non-smooth strongly nonlinear transient dynamics introduced by the BNES attachments, whose initial angle of inclination was fixed uniformly at $\phi_0 = 3.562°$. In total, several thousands of simulations are performed to adequately sample the parameter space and capture the sensitivity of the system response to variations in BNES properties.

The optimization methodology consists of a structured parametric study in which key BNES parameters, including mass, stiffness, and damping, are systematically varied. The simulation setup, execution, and post-processing are automated through a Python-based workflow that interfaces with MoorDyn. This framework enables robust management of large simulation ensembles and ensures consistent preprocessing and postprocessing across all cases. To preserve meaningful comparisons between baseline (i.e., an equal-mass cable without attached BNESs) and controlled configurations, the additional mass introduced by the BNES devices is explicitly accounted for. For all cases, the total mass of the integrated cable–BNESs system is matched to that of the baseline cable by adjusting the distributed mass density, ensuring that observed performance improvements are attributable to nonlinear energy transfer, scattering and dissipation mechanisms rather than mass-added effects. Regarding the BNESs, damping coefficients are adjusted proportionally and consistently with stiffness variations to maintain physically realistic dissipation levels. Building on the modeling framework established in (Michaloliakos et al., 2025) the optimization study is extended by introducing quantitative performance metrics that are directly relevant to subsea cable loads under VIV excitation. The selected metrics are designed to capture both, (i) the *energy exchanged between the flow and the cable*, and (ii) the resulting *cable VIV amplitudes* at representative locations along the span of the cable, thereby enabling objective comparison between competing BNES designs across different dynamic current profiles. The rationale behind considering the metric (i) is that, contrary to the linear TMD, an NES (in this case a BNES) is capable of *rapidly scattering vibration or shock energy in frequency*, typically inducing targeted (i.e., mostly irreversible) energy transfer from low-frequency cable modes to high-frequency ones; this leads to high-rate vibration attenuation of the cable response (Vakakis et al., 2022). In addition, *the strongly nonlinear energy scattering induced by the BNESs affects the fluid-cable interaction itself*, i.e., it changes the pressure distribution and its frequency content along the entire span of the cable, and this *reduces the net energy intake from the flow to the cable* (which causes VIV excitation), but also can *reduce the effective drag of the cable itself*; for an example of such effects on a sprung cylinder with an internal rotating NES undergoing VIV see (Blanchard et al., 2020). These beneficial nonlinear effects that are induced due to the strongly nonlinear action of the attached BNESs are fully verified by the numerical optimization results that follow.

In MoorDyn, the work rate (instantaneous power) associated with a force component due to fluid pressure acting on a node is obtained by the scalar product of force and nodal velocity. Accordingly, at each cable node $i$, one defines the instantaneous fluid-induced VIV power input to the cable in the cross-flow ($y$) direction, $P_{VIV,i}(t)$, as,

$$P_{VIV,i}(t) = F_{VIV,i}^{(y)}(t)\, v_i^{(y)}(t) \tag{4}$$

where $F_{VIV,i}^{(y)}(t)$ is the fluid-induced force in the $y$-direction, and $v_i^{(y)}(t)$ the cable velocity component in the $y$-direction at node $i$. Moreover, the instantaneous drag power input in the same direction and at the same node is computed as:

$$P_{drag,i}(t) = F_{drag,i}^{(y)}(t)\, v_i^{(y)}(t) \tag{5}$$

Then, the corresponding cumulative input and drag energies up to a selected time $T$ and over the specific span of the subsea cable, are computed by time integration as follows,

$$E_{VIV}(T) = \int_0^T \sum_{i \in I} P_{VIV,i}(\tau)\, d\tau \ , \quad E_{drag}(T) = \int_0^T \sum_{i \in I} P_{drag,i}(\tau)\, d\tau \tag{6}$$

where $I$ denotes the set of nodes included in the post-processing (in the present implementation, the summation is performed over all available line outputs). Numerically, the integrals in (6) are evaluated using a trapezoidal quadrature in time (Michaloliakos, 2026).

To complement the aforementioned energy-based metrics with direct measures of cable vibration severity, one computes peak-to-peak (P2P) transverse displacement amplitudes at three representative locations along the cable, namely, upstream of the first BNES location ($z \approx -5m$ – sensing location 1), between the locations of two BNES devices ($z \approx -10m$ – sensing location 2), and, lastly, downstream of the last BNES ($z \approx -19m$– sensing location 3). Let $d_y^{(m)}(t)$ denote the cross-flow cable displacement time series in the $y-$direction extracted at each sensing location $m \in \{1,2,3\}$. The corresponding P2P amplitude over a considered window $[t_a, t_b] = [0,10]$ is defined as:

$$P2P^{(m)} = \underbrace{max}_{t \in [t_a, t_b]} \ d_y^{(m)}(t) - \underbrace{min}_{t \in [t_a, t_b]} \ d_y^{(m)}(t) \ \ , \ m = 1,2,3 \tag{7}$$

These three P2P values provide a compact quantitative characterization and monitoring of the spatial distribution of cable VIV amplitudes relative to the three BNES locations and allow for the optimization procedure to target both *local* – through the P2P metrics (7), and *global* – through the energy metrics (6), VIV suppression features.

To quantify cable VIV suppression relative to a baseline design, the previous metrics are evaluated against a baseline (reference) case, which is an identical subsea cable but with no BNES attachments and with identical mass to the BNES-equipped cable (so that no added mass, or any other mass-related effects exist). For each performance metric, $\mathcal{K}$, the relative change with respect to the respective reference value $\mathcal{K}_{ref}$ is reported as, $\Delta\mathcal{K} = \mathcal{K} - \mathcal{K}_{ref}$, or $\overline{\Delta\mathcal{K}} = \frac{\mathcal{K} - \mathcal{K}_{ref}}{|\mathcal{K}_{ref}|} \times 100$. Reductions in $E_{VIV}(t_b)$, $E_{drag}(t_b)$, and the P2P amplitudes in Eqs. (6) and (7) are interpreted as improved VIV mitigation performance. Since these objectives might be inherently competing, e.g., designs that strongly suppress local cable vibrations may not simultaneously minimize the flow-to-cable energy intake everywhere along the span of the cable, the selection of designs in the computational parametric study is posed as a *multi-objective optimization problem*. Specifically, for each of the designated fluid-current case study, one identifies a Pareto-optimal set of BNES configurations by simultaneously minimizing the following objective vector $\boldsymbol{J}[\mathbb{c}]$ for the candidate BNES design $\mathbb{c}$:

$$\boldsymbol{J}[\mathbb{c}] = \left[ |E_{drag}(t_b)| \quad |E_{VIV}(t_b)| \quad P2P^{(1)} \quad P2P^{(2)} \quad P2P^{(3)} \right]^T \tag{8}$$

A candidate design $\mathbb{c}_{\boldsymbol{k}}$ is said to be *dominated* by another design $\mathbb{c}_{\boldsymbol{n}}$ (in the context of minimization) if and only if, all components of $\mathbb{c}_{\boldsymbol{k}}$ are less than, or equal to the respective components of $\mathbb{c}_{\boldsymbol{n}}$, and at least one of the components of $\mathbb{c}_{\boldsymbol{k}}$ is strictly less than the respective component of $\mathbb{c}_{\boldsymbol{n}}$; that is, design $\mathbb{c}_{\boldsymbol{n}}$ is no worse in every objective and strictly better in at least one objective compared to design $\mathbb{c}_{\boldsymbol{k}}$. *The Pareto front is then defined as the subset of designs that are not dominated by any other design in the design ensemble.*

## 4.2. Dynamic Current Profiles for VIV Excitation of the Cable

Three distinct dynamic current profiles are used as inputs for VIV excitation of the subsea cable, to systematically assess the efficacy and robustness of the set of BNESs to suppress highly non-stationary VIV of the cable under varying and highly unsteady flow conditions. For each dynamic current profile input, a total of 10,000 time-domain simulations is conducted, enabling extensive and detailed exploration of the design space and providing valuable insights with respect to the BNES parameter combinations that yield optimal VIV mitigation. All current profiles are unsteady, that is, non-uniform in depth and time-varying in magnitude, and are imposed in the global $x-$direction, thereby inducing cross-flow VIV of the subsea cable primarily in the transverse plane, as shown in Fig. 2b. The first dynamic current profile, designated as profile 1 and shown in Fig. 5a, corresponds to a moderate-intensity flow with peak velocities reaching approximately $2m/s$, with the locations of the three BNESs along the (initially) vertical span of the cable also indicated by dashed horizontal lines for reference. This current profile exhibits a relatively broad vertical extent, with significant flow penetration occurring toward the seabed. Such conditions are representative of offshore environments where VIV excitation is present but not extreme. The second current profile 2, illustrated in Fig. 5b, retains a similar depth-dependent structure, but with increased peak velocity to approximately $3m/s$. This case represents a more energetic flow regime, leading to stronger hydrodynamic forcing and more pronounced VIV response of the cable. Comparing this case to the first allows one to isolate and study the effect of increased flow intensity on the BNES vibration suppression performance while maintaining comparable vertical flow spatial distribution. It is anticipated though that, since the second current profile will induce VIV excitation of the cable at a higher energy level, the potential of the BNES (being strongly nonlinear devices) will be better harnessed. The third and final current profile 3, depicted in Fig. 5c, also reaches peak velocities of approximately $3m/s$, but is significantly more spatially confined since it is mainly localized at the upper portion of the water column and the cable. In this scenario, the strongest fluid velocities are concentrated near the free surface, with rapidly diminishing velocities at greater depths. This configuration is representative of surface-dominated current conditions and is particularly relevant for assessing the effectiveness of BNES placement strategies near the fairlead region of the subsea cable. Together, these three current profiles span a representative range of realistic offshore conditions, enabling comprehensive evaluation of the efficacy and robustness of the performance of the attached BNESs across highly unsteady and strongly varying in space and time dynamic currents.

To systematically explore the high-dimensional design space of the three BNES parameters, a large ensemble of simulation cases is generated using Latin Hypercube Sampling (LHS). This approach enables efficient and space-filling coverage of the parameter domain while keeping the total number of simulations computationally tractable. For each current profile 10,000 distinct MoorDyn simulations were performed, each for a unique combination of system parameters of the BNESs, while all other cable, hydrodynamic, and environmental parameters are held fixed as defined previously. Each candidate design consists of three BNES masses, $M_j$ and three axial stiffness values, $EA_j$ (of the inclined elastic chords-in-tension – see Fig.3a). The initial angle of inclination is fixed uniformly at $\phi_0 = 3.562°$. Accordingly, the $(6 \times 1)$ design vector for a given case is written as,

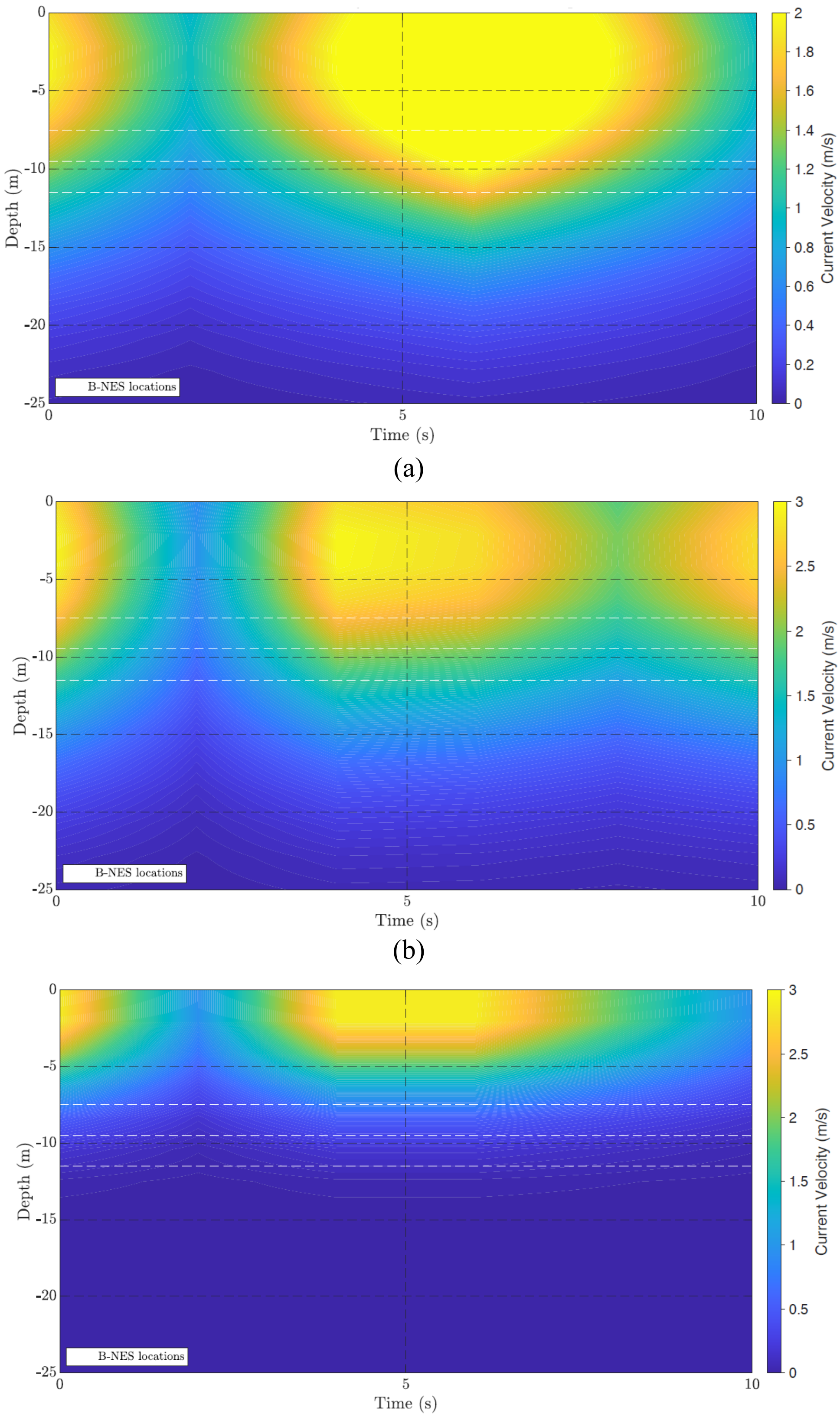

(a)

(b)

(c)

Figure 5. Unsteady, time- and space-varying dynamic current profiles for the multi-objective optimization study: (a) Profile 1, peak velocity $\sim 2m/s$ (moderate VIV), (b) profile 2, peak velocity $\sim 3m/s$ (more intense VIV), and (c) profile 3, peak velocity $\sim 3m/s$ localized near free surface (surface-dominated VIV excitation); dashed lines show BNES locations.

$$\mathbb{C} = [M_1 \quad M_2 \quad M_3 \quad EA_1 \quad EA_2 \quad EA_3]^T \tag{9}$$

where the masses are sampled uniformly in the interval $M_j \in [3,5]Kg$, and the stiffnesses are sampled logarithmically, $log_{10}EA_j \in [6,10]N$ , i.e., $EA_j \in [10^6, 10^{10}]N$ . Sampling the stiffness on a logarithmic scale is essential, as the nonlinear response of the integrated cable–BNES system exhibits sensitivity across several orders of magnitude in axial stiffness. To prioritize exploration of the most influential parameters, the sampling strategy is deliberately biased toward richer variation in stiffness while limiting redundancy in the mass combinations. For each of the three dynamic current profiles, a full set of 10,000 LHS samples is generated for the three stiffness parameters, ensuring that each simulation corresponds to a unique stiffness triplet. In contrast, a smaller set of distinct mass triplets is generated first and then tiled across the full ensemble of cases. This strategy preserves variability in mass while allowing for the computational effort to be focused on resolving the strongly nonlinear dependence of the system response on stiffness. For each sampled stiffness value, it was assumed that a linear internal damping coefficient of each BNES exists which is in proportion to the stiffness of each elastic cord, so to maintain consistent classical damping distribution across the entire design space. Specifically, the linear damping coefficient of the BNES is defined as, $c_j = max\left(\frac{EA_j}{10^6}, 300\right)$, which enforces a minimum damping threshold while increasing dissipation for stiffer configurations. This scaling prevents unrealistically underdamped behavior at high stiffness and improves numerical stability of the time-domain simulations.

All candidate designs are constructed by modifying a single baseline MoorDyn input file. For each case, only the BNES-related parameters are overwritten: The point masses associated with the BNES nodes and the axial stiffness and damping properties of the cords-in-tension. Each modified input file is written to disk and assigned to a unique case identifier, with the sampled parameter values encoded in the filename to ensure full traceability between simulation inputs and outputs. This LHS-based case generation provides a statistically well-distributed and computationally efficient foundation for the subsequent optimization analysis. By densely sampling stiffness over a wide dynamic range while retaining controlled mass variability, the resulting ensemble of simulations enables robust identification of the three BNES configurations that consistently mitigate VIV-induced vibrations across the dynamic current scenarios considered in this study.

Starting with dynamic current profile 1 (see Fig. 5a), a way to assess the VIV mitigation performance of the BNESs using the cumulative energy-based metrics (6) and peak-to-peak displacement measures (7), both assembled in the objective vector (8). For each flow current profile, the optimization ensemble consists of computational realizations of 10,000 candidate BNES configurations generated via Latin Hypercube Sampling, with each design evaluated through a full time-domain MoorDyn simulation. The resulting performance data are summarized using normalized scatter plots that visualize the trade-offs between reducing, (i) the flow-to-cable energy net intake, and (ii) the transverse VIV cable amplitudes at multiple sensor locations along the cable (as described previously).

## 4.3. Optimization Results and Nonlinear Dynamics Governing VIV Cable Mitigation

Fig. 6 presents the resulting normalized performance maps obtained from the BNES optimization study for dynamic current profile 1. The subplots in Fig. 6a–f provide an assessment of the BNES performance in mitigating VIV of the cable, both in terms of reducing cable vibration amplitudes at the sensor locations and reducing the cumulative energy intake and drag amplification from the flow. The vertical axis in each subplot shows the normalized P2P transverse VIV amplitude of the cable evaluated at each of the three representative sensing locations along the cable; while the horizontal axis reports either the normalized cumulative drag energy, $E_{drag}(t_b)$, or the normalized cumulative VIV input energy, $E_{VIV}(t_b)$, both evaluated at the terminal time $t_b = 10s$. Each point represents one realization of the 10,000 sampled BNES designs, thereby providing a direct visualization of the multi-objective trade-offs inherent to the VIV mitigation problem. For each current profile, a representative *Pareto-optimal* configuration is selected from the nondominated set obtained that minimizes the magnitude of the objective vector normalized in each dimension in (8). The selected Pareto-optimal BNES design is indicated by a red marker in Fig. 6, and the optimized system parameters for the three BNESs are listed in Table 1. The respective Pareto optimization for dynamic current profile 2 (Fig. 5b), which is similar to profile 1 but of higher VIV intensity, is given in the Appendix, and the results are similar to Fig. 6. Taken together, these results demonstrate that the developed data-driven optimization framework consistently identifies BNES configurations that simultaneously reduce global energy input from the flow, reduce drag energy and drag amplification of the cable in VIV, and suppress local vibration amplitudes along the cable, even as the current excitation intensity increases from current case 1 to current case 2.

These results reveal an unexpected (and perhaps counter-intuitive) benefit of the proposed strongly nonlinear VIV mitigation design: Apart from the significant and consistent reduction in the VIV amplitudes of the cable, *the action of the BNESs significantly reduces the energy intake to the cable from the flow, yielding similar reduction in the corresponding cumulative VIV-induced drag energy and amplification.* The physical reason controlling the reduction of these energy measures lies in the capacity of the BNESs to rapidly scatter VIV energy from low-to-high frequency cable modes – see also (Michaloliakos et al., 2025), which, as a result, changes fundamentally the fluid-cable interaction, e.g., pressure distribution exerted by the flow along the entire span of the cable. This changes the effective mechanical impedance of the cable and its capacity to exchange energy with the surrounding fluid, which, in turn, affects the energy intake in the cable during VIV, as well as the corresponding drag amplification of the cable. *These results indicate that strongly nonlinear attachments can be optimally designed to significantly reduce the energy intake from the surrounding flow and the drag experienced by subsea structures,* which adds a completely new dimension in the VIV mitigation problem considered herein. Lastly, it is noted that, while similar reductions in energy intake in structures from their environment have been reported in other contexts (Blanchard et al., 2020; Gzal et al., 2023), to the authors' knowledge this is the first time that this nonlinearity-induced energy benefit has been reported in VIVs of subsea cables.

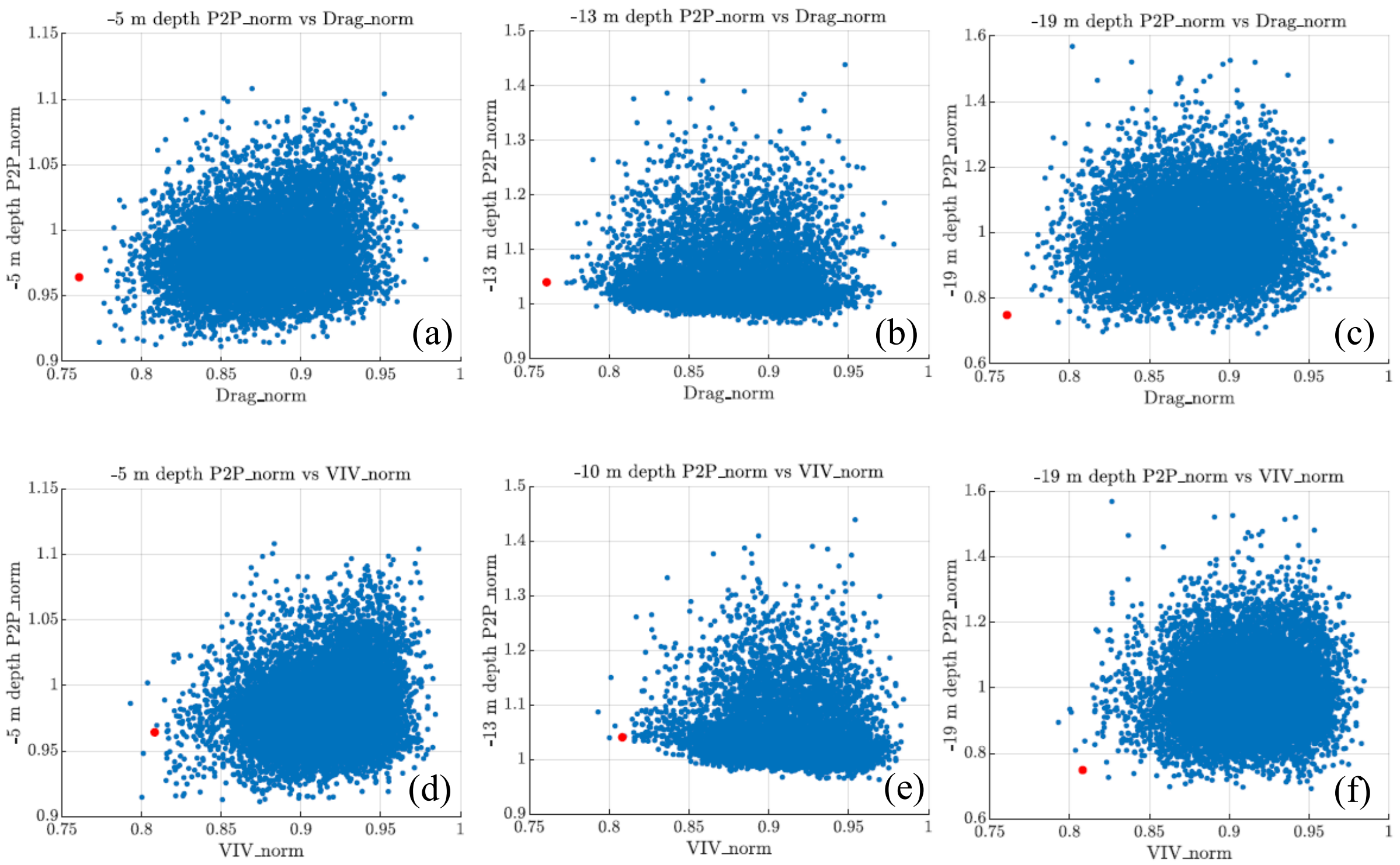

Figure 6. Normalized performance maps for current profile 1: (a-c) P2P transverse VIV amplitude versus cumulative VIV-induced drag energy to the cable at the three sensing locations on the cable; (d–f) P2P transverse VIV amplitude versus cumulative VIV-induced energy intake to the cable (10,000 sampled BNES designs shown in each plot, and red marker denotes the selected Pareto-optimal BNES configuration).

Table 1. Current profile 1: Optimal parameters of the BNESs (see Fig. 2a).

| BNES | $EA_i$ (N) | $d_i$ (Pa-s) | $\varphi_0$ (deg) | Mass (kg) |
|---|---|---|---|---|
| 1. Top | $1.39\times10^6$ | $3.00\times10^2$ | 3.562 | 4.81 |
| 2. Intermediate | $3.81\times10^9$ | $3.81\times10^3$ | 3.562 | 4.09 |
| 3. Bottom | $3.62\times10^7$ | $3.00\times10^2$ | 3.562 | 4.31 |

Apart from current profile 3 which will be discussed in more detail below, comparing the optimal configurations against the baseline cable dynamics in terms of VIV-induced cumulative input and drag energies reveals a consistent and physically intuitive trend. This is shown in Fig. 7 for current profiles 1 and 2. As the current velocity (excitation) intensity increases, the relative energy reduction achieved by the action of the three BNESs becomes more pronounced. Indeed, for the lower-intensity current case 1 (Figs. 7a,b), the optimized BNES configurations yield a reduction of approximately 13% in cumulative VIV energy intake, and nearly 15% in cumulative drag energy (and respective drag amplification). For comparison, for the higher-intensity current case 2 (Fig. 7c,d), the corresponding reductions increase to approximately 19% for VIV energy intake, and nearly 24% for drag energy.

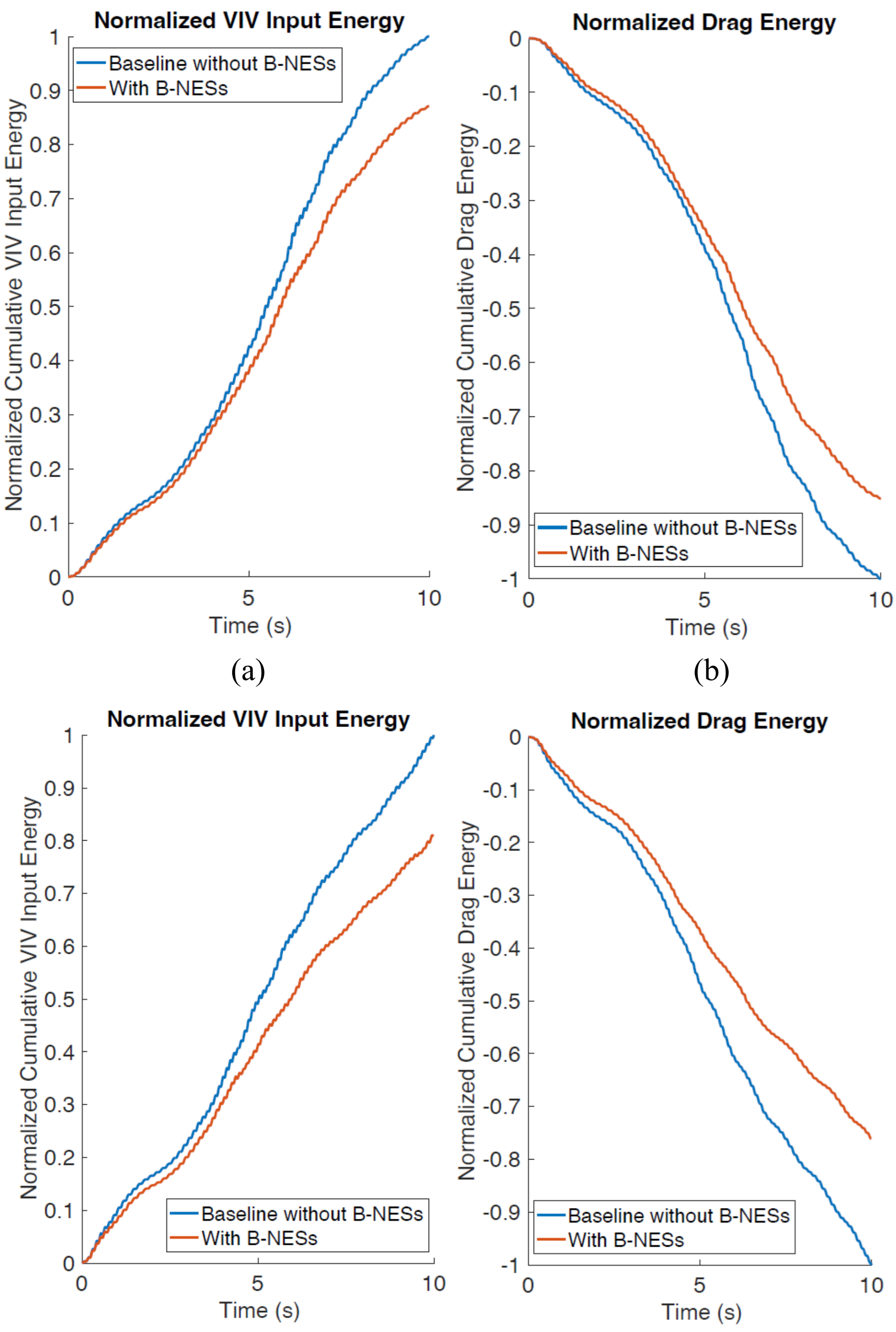

(c) (d)

Figure 7. Comparisons of baseline and optimized BNES-equipped subsea cables undergoing VIV: Cumulative normalized VIV-induced input energy to the cable (a), and corresponding hydrodynamic drag energy (b) for the less intense current profile 1 (Fig. 5a), and respective energy metrics (c,d) for the more intense current profile 2 (Fig. 5b).

Such enhanced performance at higher excitation levels is expected, since the strongly nonlinear bi-stable attachments are inherently energy-dependent and broadband in their operation (Vakakis et al., 2008, 2022; Michaloliakos et al., 2025), in contrast to linear devices, e.g., TMDs, whose performance is not affected by the intensity of the excitation, and their effectiveness is typically limited to narrow frequency regimes (Ormondroyd and Den Hartog, 1928). It follows that higher incoming energy levels activate the nonlinear effects of the BNESs more strongly, enabling increased VIV energy scattering in the modal space of the cable, enhanced targeted VIV energy transfer to, and local energy dissipation at the BNESs; this results in superior VIV mitigation in the cable, even subject to more severe and highly unsteady VIV conditions.

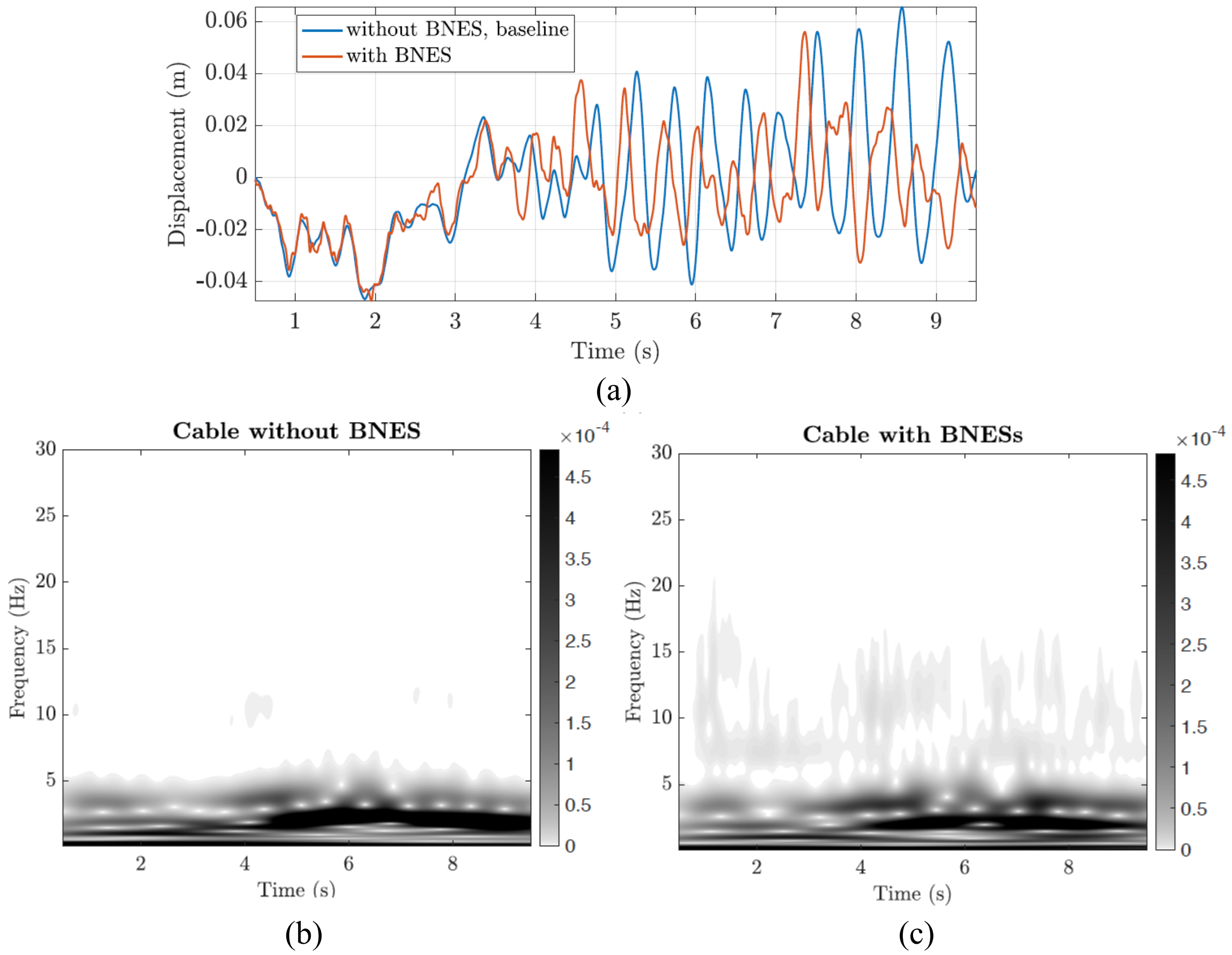

Figure 8. VIV cable mitigation for current profile 1: (a) Transverse displacement time histories of the baseline and BNES-equipped cables at a location downstream of the three BNES attachments; Wavelet transform spectra of, (b) the baseline cable displacement showing energy concentration at low frequencies associated with VIV, and (c) the BNES-equipped cable displacement, illustrating nonlinear low-to-high frequency energy scattering in the cable itself.

The corresponding P2P amplitude changes of the optimized designs relative to the baseline cable at the three sensing locations ($z \approx -5$, $-10$ and $-19$ m) are −4.1%, −1.3% and −14.6% for current profile 1, −3.6%, +4.1% and −25.1% for current profile 2, and −6.7%, −7.0% and −31.0% for current profile 3. The largest reductions occur downstream of the BNESs (sensing location 3), reaching 15–31%. At the intermediate location (sensing location 2), the selected design for current profile 2 exhibits a marginal increase of 4.1%, which reflects the multi-objective nature of the optimization: The selected Pareto-optimal design trades a small local amplitude increase between two BNESs for substantial reductions in the global energy metrics and in the downstream amplitude. Since fatigue damage scales with a high power $m$ of the stress range (typically $m \approx 3$–$5$ for the metallic components of power cables), an amplitude reduction by a fraction $r$ reduces the fatigue damage per cycle by the factor $(1 - r)^m$; for instance, the 15–31% downstream reductions correspond, under the usual assumption of stress proportional to the vibration amplitude, to reductions of the damage per cycle of approximately 38–84%. A complete fatigue assessment, based on rainflow counting over long-term current statistics, is left for future work.

The energy dependence of BNES performance also implies a bounded working (efficiency) range (Manevitch et al., 2014; Romeo et al., 2015). Below a critical energy the BNES mass performs in-well oscillations and behaves essentially as a detuned linear absorber; above it, cross-well (and possibly chaotic) motions activate strongly nonlinear, broadband energy transfers; and at still higher energies the nonlinear attachment is overpowered (in essence, due to high frequency oscillation it behaves as a rigid body attachment) and its efficiency decreases. The increase of the relative energy reductions from current profile 1 to current profile 2 (Fig. 7) is consistent with BNES operation inside this working range; the working range of the optimized BNESs is examined systematically in Section 4.5.

Following the optimization study, the consistency of the obtained results is verified, in consideration of previously reported findings on BNES-equipped subsea cables subject to platform-induced vibration but not undergoing VIV (Michaloliakos et al., 2025). To this end, one considers the displacement time histories and time–frequency responses of the subsea cable equipped with optimized BNESs and subject to dynamic current profile 1. Specifically, focus is given to the transverse displacement response at a representative cable location downstream of the three BNES attachments, where the cumulative effects of nonlinear energy exchanges between the fluid and the cable is expected to be most pronounced.

In Fig. 8a a comparison is given of the transverse displacement time series of the cable with and without the optimized BNES attached. The nonlinear action of the BNESs leads to clear reduction of the cable vibration amplitude, which is accompanied with a modulation of the response indicative of nonlinear energy scattering of energy from low-to-high frequencies within the cable itself. This behavior is fully consistent with the low-to-high VIV nonlinear targeted energy transfer that is induced by the BNESs, whereby energy is extracted from the dominant (and higher-amplitude) low-frequency VIV cable modes and scattered mostly irreversibly to (lower-amplitude) higher-frequency modes of the cable; indeed, as VIV energy is transferred from low-to-high frequencies the overall amplitude of the cable vibration reduces accordingly, while the vibration energy is much more efficiently dissipated by the higher frequency modes that possess more effective dissipative capacity.

To further elucidate this underlying nonlinear energy transfer mechanism, Figs. 8b and 8c show the corresponding wavelet transform spectra of the cable displacement for the baseline cable and the cable equipped with optimized BNESs respectively. In the baseline configuration, the vibration energy of the cable is concentrated within a narrow low-frequency band associated with the low-frequency cable modes that excite the vortex-induced vibrations. A different result appears in the plot of Fig. 8c, where due to the action of the optimized BNESs, the wavelet spectrum reveals pronounced scattering of VIV energy toward higher frequencies, together with broadband energy spreading in time. This energy scattering from low-to-high frequencies is a defining feature of strongly nonlinear NES attachments and physically explains the observed reduction in the vibration amplitudes and energy metrics for the BNES-equipped subsea cable.

More physical insight into the nonlinear dynamics of the optimized BNESs (with system parameters listed in Table 1) when mitigating VIV of the cable, is gained by examining the relative transient motion of each BNES mass with respect to the motion of the points of the cable where it is attached. Figs. 9a-c depict the relative displacement time histories for the three optimized BNESs for VIV induced by current profile 1. The responses exhibit intermittent large-amplitude oscillations, characteristic of cross-well motions in bi-stable systems. These cross-well oscillations are a key mechanism enabling rapid low-to-high frequency scattering of VIV energy, yielding targeted VIV energy transfers to the higher frequency modes of the cable itself. This high-frequency energy scattering is better evidenced in Figs. 9d-f which shows the corresponding wavelet transform spectra of the relative BNES responses. These time-frequency representations demonstrate that each optimized BNES operates over a broad frequency range, absorbing VIV energy from the low-frequency cable motion and redistributing (scattering) it towards higher frequencies. This broadband behavior confirms that the BNESs act as passive, strongly nonlinear energy scatterers, rather than narrowband absorbers, and highlights their inherent passive tunability. In fact, as discussed in (Vakakis et al., 2008), since the NESs lack a preferential (preset) resonance frequency, they have the capacity to tune their resonance frequencies depending on the energy and frequency content of the applied excitations, and, hence, engage in resonance captures (Arnold, 1988) with different modes of the structures to which they are attached (in this case the cable). This feature renders the BNESs ideal passive mitigation devices for highly transient and non-stationary VIV that are induced by unsteady dynamic currents.

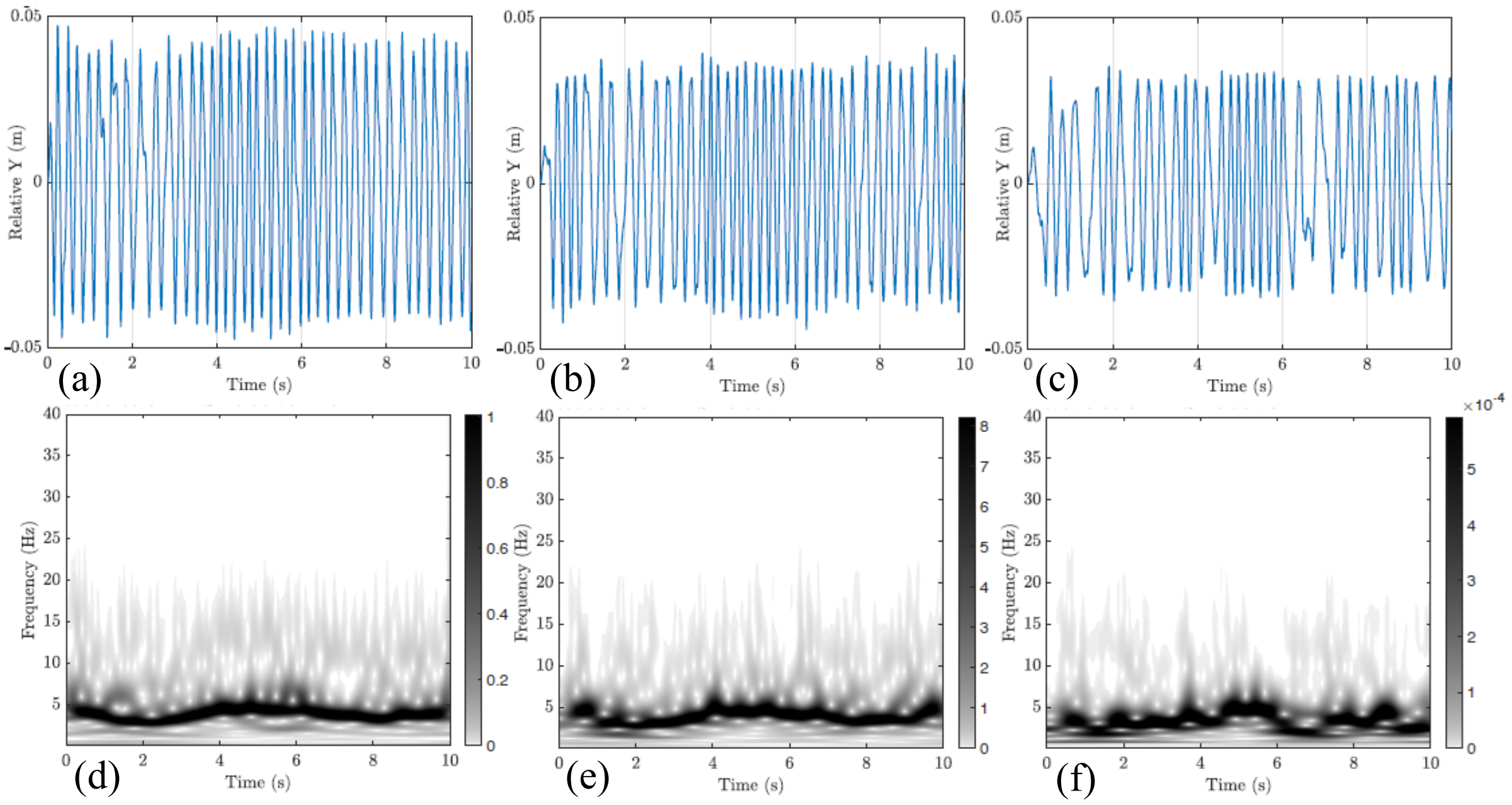

Figure 9. Nonlinear dynamics of the three optimized BNESs for dynamic current profile 1: (a–c) Relative displacement time histories of the BNESs masses with respect to the cable showing intermittent cross-well oscillations; (d–f) corresponding wavelet transform spectra of the relative responses in (a-c) illustrating broadband, nonlinear VIV energy scattering from low-to-high frequencies.

The intermittent large-amplitude relative responses of Figs. 9a–c, together with the broadband wavelet spectra of Figs. 8c and 9d–f, are consistent with chaotic cross-well dynamics, which have been documented for a linear oscillator coupled to a bi-stable attachment together with the energy thresholds separating cross-well, regular in-well and chaotic cross-well responses (Manevitch et al., 2014; Romeo et al., 2015). The broadband scattering observed in the present work is, at least in part, a manifestation of such chaotic cross-well motion, which spreads the energy of the BNESs, and through them of the cable, over a broad frequency band and thereby favours dissipation. On the other hand, chaos implies sensitivity to initial conditions and to the particular realization of the current, as well as intermittent large excursions of the BNES masses; this sensitivity is quantified in Section 4.5, where it is shown to be confined only to the BNES-equipped cable and to affect the integral energy metrics only marginally. It is noted, moreover, that the BNES strokes are bounded by the hardening of the cords-in-tension, and that the integral energy metrics (6) are inherently less sensitive to such variability than point-wise amplitudes. A rigorous characterization of the chaotic dynamics (e.g., through Lyapunov exponents) of the present high-dimensional, non-autonomous fluid–cable–BNES system, and of its implications for fatigue-load statistics, is left for future work.

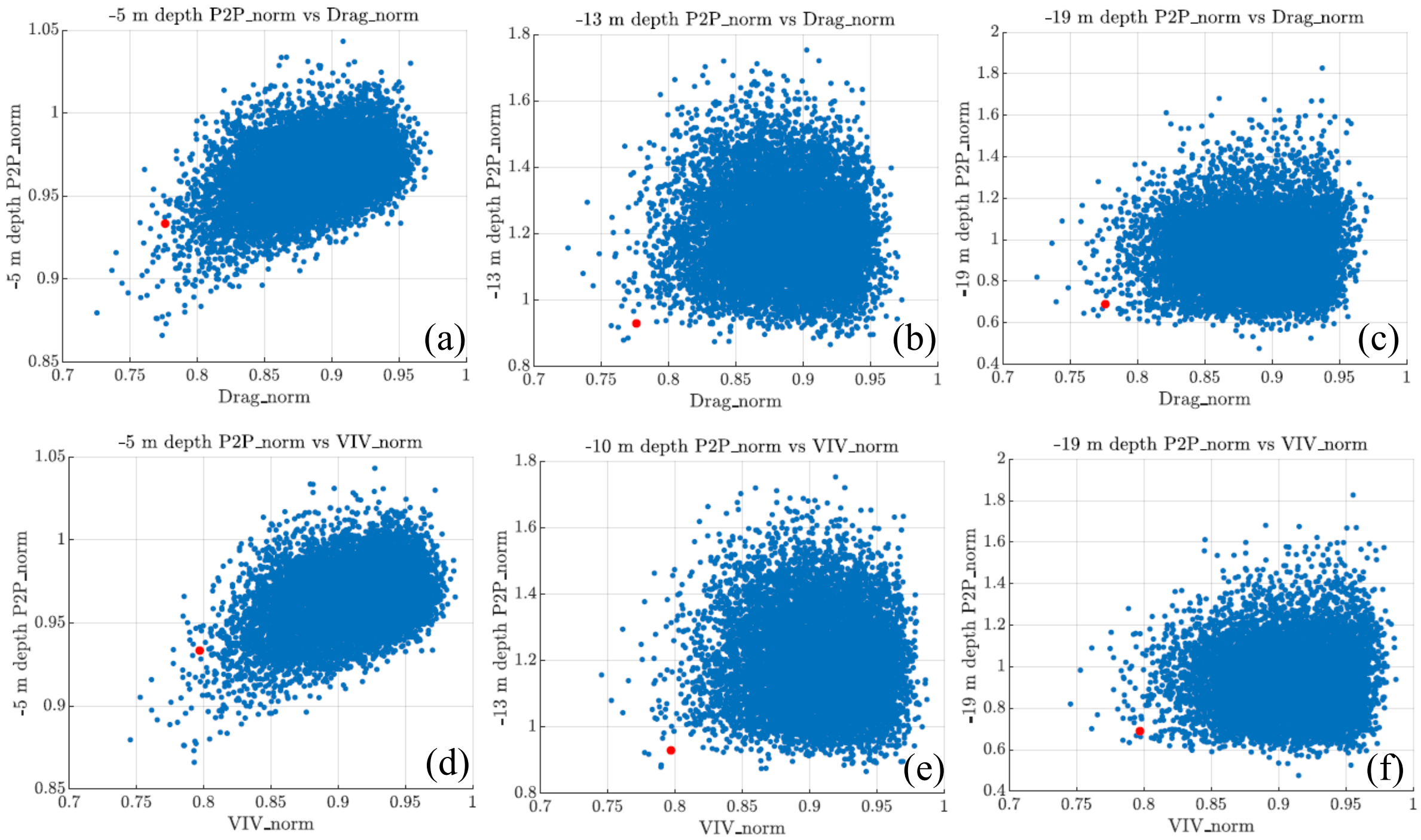

Figure 10. Normalized performance maps for current profile 3: (a-c) P2P transverse VIV amplitude versus cumulative VIV-induced drag energy to the cable at the three sensing locations on the cable; (d–f) P2P transverse VIV amplitude versus cumulative VIV-induced energy intake to the cable (10,000 sampled BNES designs shown in each plot, and red marker denotes the selected Pareto-optimal BNES configuration).

Lastly, the third dynamic current profile is considered, characterized by a high peak velocity concentrated near the free surface, as shown in Fig. 5c. This profile represents a practically important scenario for subsea power cables, since surface-intensified currents commonly occur in offshore environments and are particularly relevant when combined with platform motions that may induce heave-driven VIV to the subsea cable. Following the same optimization framework adopted for dynamic current profiles 1 and 2, we populate the design space of the three BNESs using the previously defined performance metrics. Figure 10 presents the resulting normalized performance maps, where peak-to-peak transverse displacement is reported against cumulative energy input metrics at three representative locations along the cable. As previously, each point corresponds to one of the 10,000 sampled BNES designs, while the red marker denotes the selected Pareto-optimal B-NES configuration obtained by minimizing the combined objectives of cumulative VIV energy, cumulative drag energy, and local cable vibration amplitudes. Consistent with the earlier cases, the optimized BNESs yield simultaneous reductions in both VIV-induced and drag-induced cumulative energies of the cable.

These reductions are depicted in Fig. 11. The surface-localized (and shock-like) nature of the excitation enhances the effectiveness of the strongly nonlinear attachments, since the BNESs, lacking preferential resonance frequencies and with inherent dynamics that are tunable to the applied load, are best suited to address such types of highly non-stationary, highly energetic, transient excitations (Vakakis et al., 2008). This is precisely the case of current profile 3, where energy is localized closer to regions of elevated flow velocity (in the upper part of the cable close to the platform), and there is where VIV is most profound.

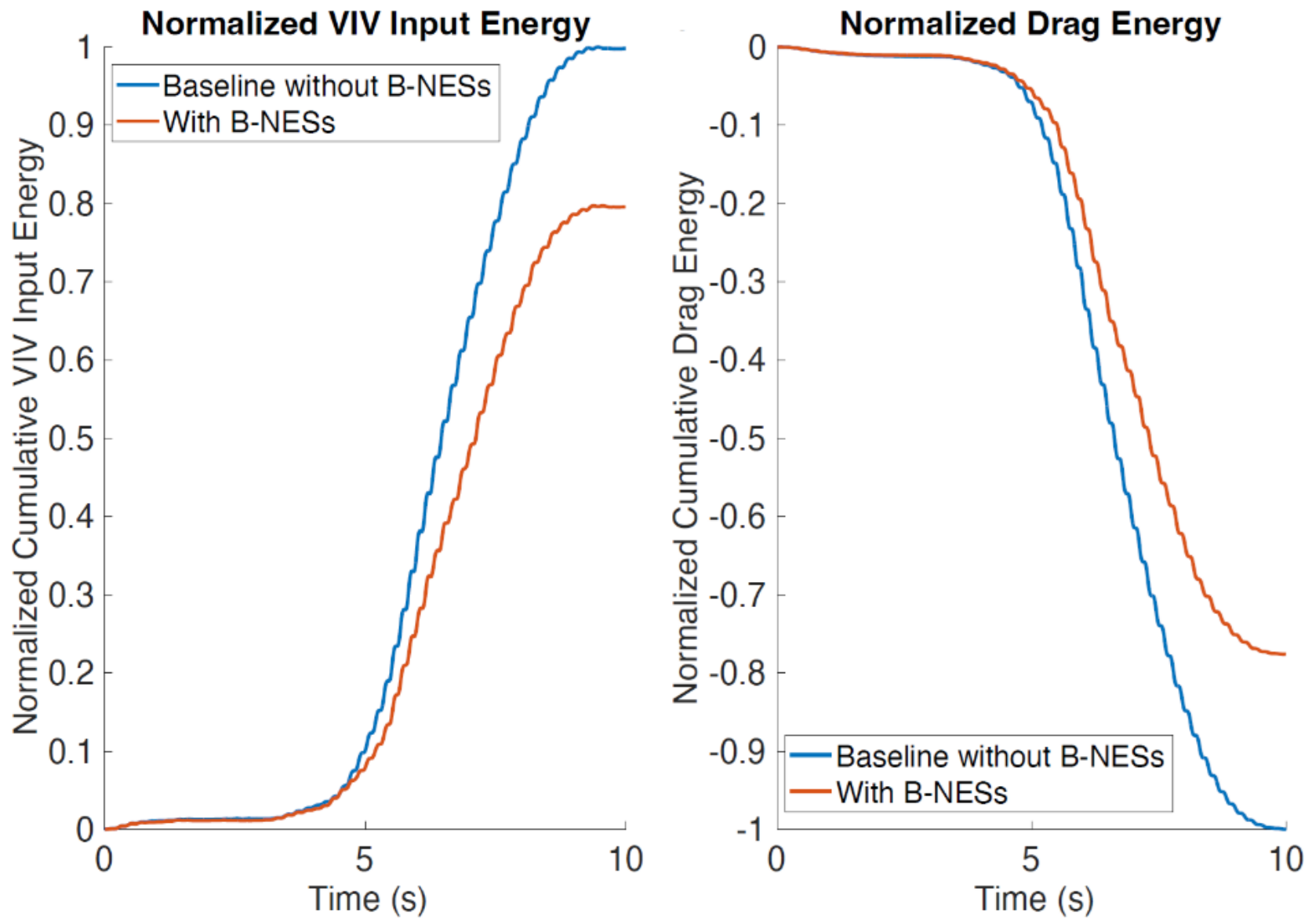

(a) (b)

Figure 11. Comparison of baseline and optimized BNES-equipped subsea cables undergoing VIV for current profile 3 (Fig. 5c): (a) Cumulative normalized VIV-induced input energy to the cable, and (b) corresponding hydrodynamic drag energy.

The reported results confirm that strongly nonlinear passive devices, such as the BNESs considered in this study, are particularly well suited for mitigating VIV in scenarios dominated by near-surface forcing, where the VIV-induced loads are of high-intensity, strongly localized, highly non-stationary and of multi-frequency content (broadband). The system parameters of the optimized BNESs are listed in Table 2.

Table 2. Current profile 3: Optimal parameters of the BNESs (see Fig. 2a).

| BNES | $EA_i$ (N) | $d_i$ (Pa-s) | φ0 (deg) | Mass (kg) |
|---|---|---|---|---|
| 1. Top | $1.03\times10^7$ | $3.00\times10^2$ | 3.562 | 4.78 |
| 2. Intermediate | $2.45\times10^6$ | $3.00\times10^2$ | 3.562 | 4.55 |
| 3. Bottom | $3.01\times10^9$ | $3.01\times10^3$ | 3.562 | 4.99 |

To further elucidate the nonlinear physical mechanisms that govern the observed performance gains due to the actions of the optimized BNESs, the instantaneous power input to the cable due to VIV forcing is considered. This is computed by the inner product of the VIV force exerted to the cable, and the local cable velocity at the point (node) of application of that force. Fig. 12a compares the VIV power time histories for the baseline and optimized BNES-equipped cable subject to current profile 3. It is noted that the nonlinear dynamics of the optimized BNESs introduces a marked reduction in peak VIV power levels, together with a redistribution of power over shorter time scales, indicating a direct reduction in cable velocity response and, corresponding reduction in drag amplification caused by the VIV of the cable.

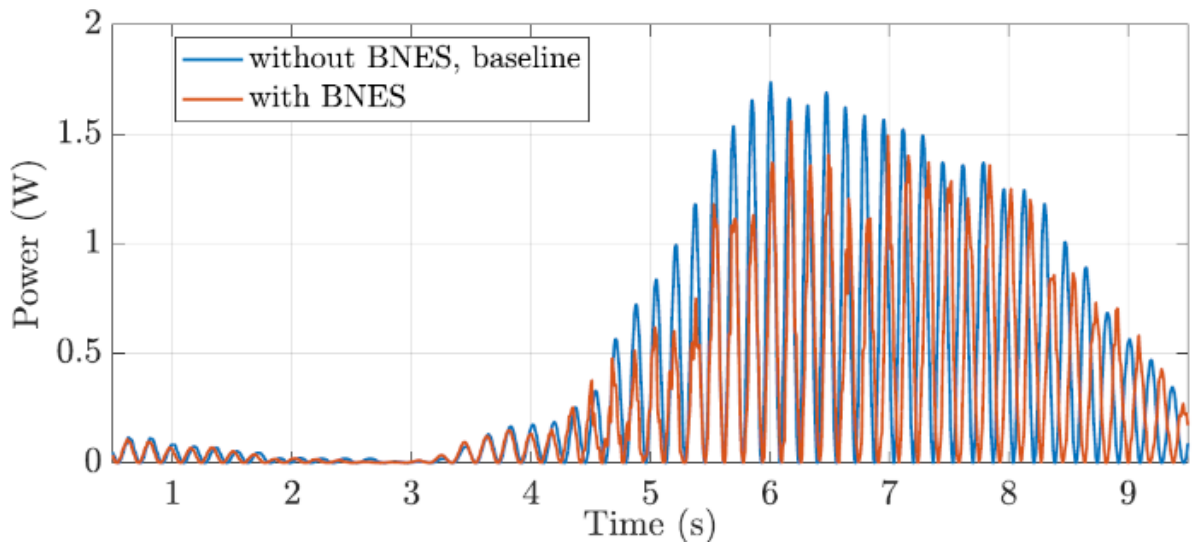

(a)

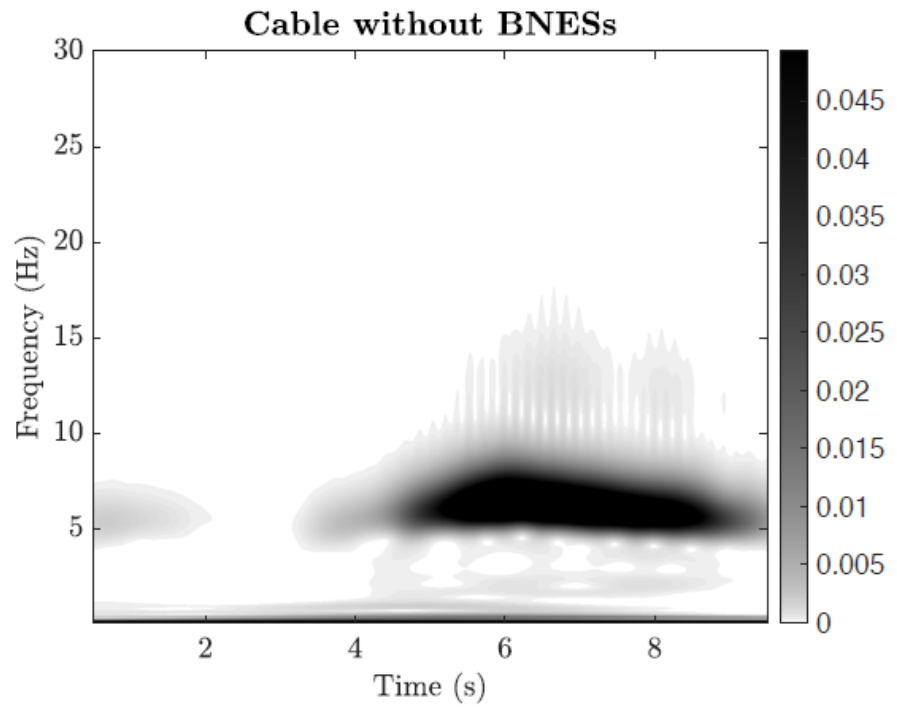

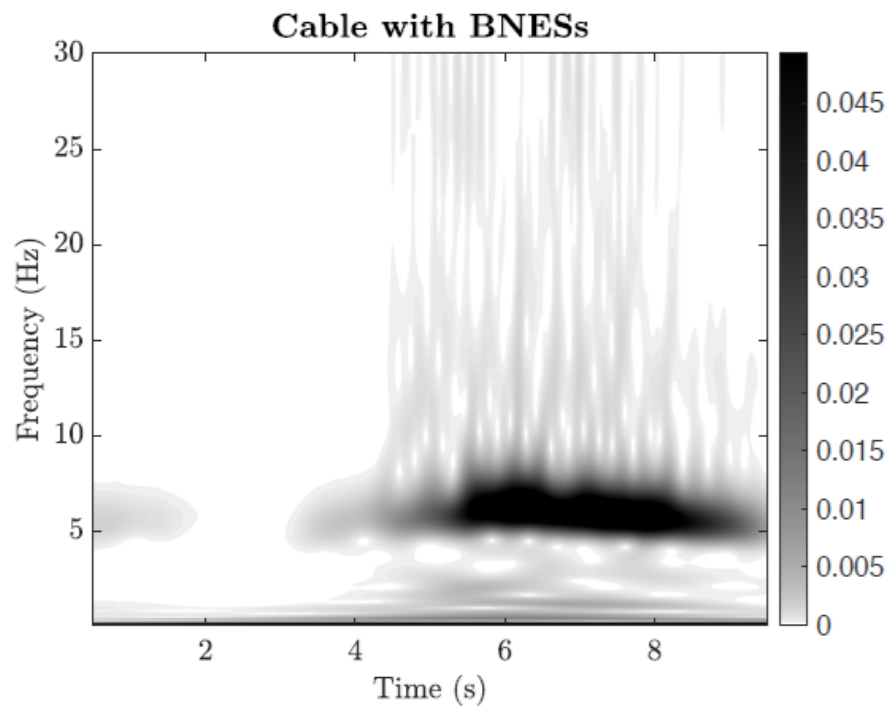

(b) (c)

Figure 12. Power-based analysis for dynamic current profile 3: (a) Time histories of instantaneous VIV power input to the baseline cable and the cable with optimized BNESs, (b,c) corresponding wavelet transform spectra of the power signals in (a), highlighting low-to-high VIV energy scattering by the actions of the BNESs.

Additional insight is provided by the time–frequency wavelet analysis of the power signals shown in Figs. 12b,c. In the baseline configuration, the power spectrum is dominated by low-frequency components associated with the large-amplitude VIV response of the cable. Due to the action of the optimized BNESs, the wavelet spectrum of the VIV power input for the BNES-equipped cable reveals a clear targeted VIV energy transfer from low-to-high frequencies. This is a direct manifestation of the nonlinear energy scattering (redistribution) of the modal energies of the cable induced by the BNESs, whereby energy extracted from the dominant low-frequency VIV of the cable is transferred (directed) to higher-frequency components where VIV is more readily dissipated by hydrodynamic damping and the inherent dissipative capacity of the cable itself; at the same time, as energy is transferred from low-to-high frequencies the amplitude of the cable vibration reduces accordingly, providing an additional benefit of this strongly nonlinear mitigation design.

The specific benefit of bi-stability, as opposed to a monostable nonlinear attachment, was assessed in previous work on the same cable–BNES configuration (Michaloliakos et al., 2025; Michaloliakos, 2026). There, equal-mass monostable NESs with purely cubic stiffness, obtained from the same device by setting $\phi_0 = 0$, were clearly outperformed by the optimized BNESs, owing to the additional low-to-high frequency energy scattering enabled by cross-well transitions. In the same previous work, the robustness of the BNES to shifts of the cable natural frequencies caused by biofouling was demonstrated without any re-tuning; clearly such passive tunability in the dynamics cannot be achieved by a linear TMD whose effectiveness relies on matching a fixed natural frequency, and any detuning results in drastically diminished performance (Ormondroyd and Den Hartog, 1928; Marty et al., 2021). These findings motivate the use of BNESs for the strongly non-stationary, multi-frequency VIV considered herein; a direct comparison with equal-mass monostable NESs under the present VIV conditions is reported in Section 4.5. In the next Section we quantify the VIV energy redistribution in the modal space of the cable due to the actions of the BNESs.

### 4.4. Modal-level Evidence of Low-to-High Energy Transfers within the Cable

The wavelet spectra discussed above are point-wise, so that, by themselves, they demonstrate a redistribution of energy in frequency rather than among the structural modes of the cable; higher-frequency content at a point could, in principle, also arise from local super-harmonics of the strongly nonlinear BNES forces. To provide modal-level evidence, the natural frequencies of the baseline cable (without BNESs, in still water) were first identified from a free-decay MoorDyn simulation, initiated by a short (0.1 s) raised-cosine lateral displacement pulse of 0.02 m at the fairlead and followed by 30 s of free vibration; the cross-flow displacement spectra of all cable nodes were summed and the resonance peaks identified. The resulting natural frequencies, listed in Table 3, are well separated.

Table 3. Natural frequencies of the baseline cable (still water) identified from free decay.

| Mode | $f_n$ (Hz) | Mode | $f_n$ (Hz) | Mode | $f_n$ (Hz) |
|---|---|---|---|---|---|
| 1 | 0.75 | 7 | 9.44 | 13 | 17.50 |
| 2 | 1.88 | 8 | 10.44 | 14 | 20.44 |
| 3 | 2.94 | 9 | 12.38 | 15 | 21.38 |
| 4 | 3.44 | 10 | 13.63 | 16 | 22.19 |
| 5 | 5.38 | 11 | 15.19 | 17 | 22.75 |
| 6 | 7.50 | 12 | 16.06 | 18 | 24.56 |

Since the natural frequencies are well separated, the response within a frequency band around each natural frequency is dominated by the corresponding mode. For the optimized design of each current profile and the corresponding baseline, the simulations were repeated with a finer output sampling (they reproduce the original P2P metrics to within 1%), and the cross-flow velocity power within each modal band was computed and summed over all nodes along the entire span of the cable, so that the result is not tied to a single measurement point. The results are shown in Fig. 13a–c. In the baseline cable more than 98% of the VIV energy is captured by the low-order modes (0.75–5.38 Hz) that are synchronized with vortex shedding. Due to the action of the BNESs, the energy in these VIV-dominated modes decreases by 17%, 26% and 14% for current profiles 1, 2 and 3, respectively, whereas the energy in the higher-order modes (7.5–25 Hz), which are not directly excited by vortex shedding, increases by factors of approximately 7, 7 and 4, with individual modes gaining up to two orders of magnitude. Accordingly, the share of the total cable vibration energy carried by the higher-order modes rises from 0.7% to 5.5%, from 0.6% to 5.5% and from 1.8% to 8.3% for the three current profiles 1,2 and 3, respectively, while the total span-integrated vibration energy decreases by 13%, 22% and 7%.

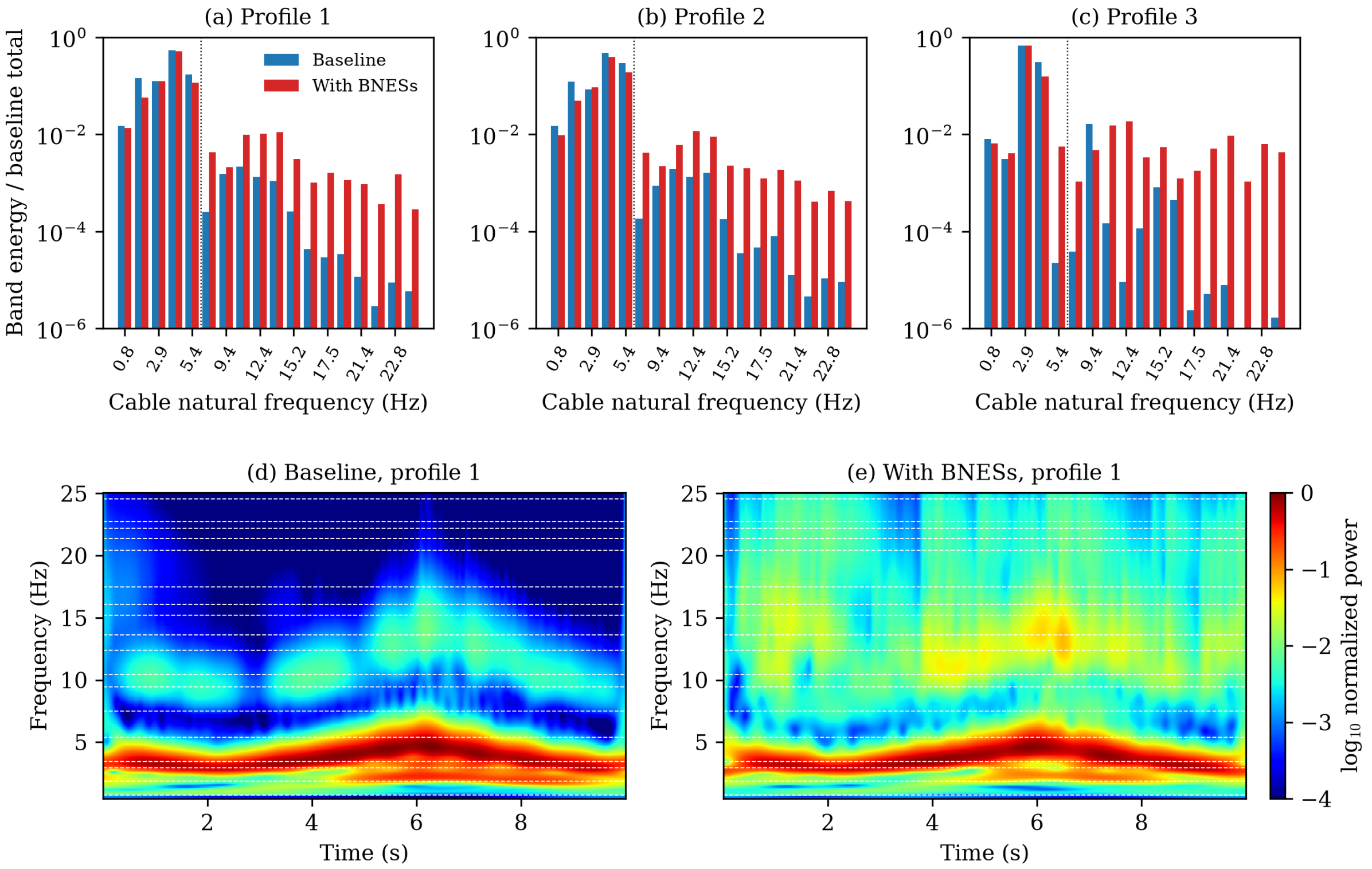

Figure 13. Modal-level analysis of the VIV energy transfer: (a–c) Span-integrated cross-flow velocity power within the band of each natural frequency of the cable (Table 3) for the baseline and the optimized BNES-equipped cable, normalized by the baseline total, for current profiles 1–3 (the dotted vertical line separates the VIV-dominated modes from the higher-order modes); (d,e) span-summed wavelet power of the cross-flow velocity for current profile 1, (d) baseline cable and (e) cable with BNESs; dashed lines denote the natural frequencies of the cable.

The span-summed wavelet spectra of Figs. 13d,e further show that the transferred energy concentrates at the natural frequencies of the cable (dashed lines) and is present over the whole span rather than only near the BNESs, indicating that it populates cable modes rather than consisting solely of locally generated super-harmonics. These results provide direct, span-integrated evidence that the BNESs transfer VIV energy from the dominant low-order cable modes to higher-order modes, in analogy with the intermodal targeted energy transfer (IMTET) mechanism reported in Gzal et al. (2023) for a seismically excited building model. The higher-frequency modes exhibit much smaller displacement amplitudes for a given energy and dissipate energy more efficiently through hydrodynamic and structural damping.

### 4.5. Working Range, Sensitivity and Comparison with a Monostable NES

To examine the working range of the optimized BNESs, the design optimized for current profile 2 (Table A.1 in the Appendix) was kept fixed, and the amplitude of the dynamic current of profile 2 was scaled to peak velocities from 1.5 to 3.5 m/s, i.e., over a range that encloses the intensities of current profiles 1 (approximately 2 m/s) and 2 (approximately 3 m/s), with the baseline cable recomputed at each intensity. The resulting reductions are shown in Fig. 14a, together with the fraction of time during which the cords of each BNES are in tension (Fig. 14b), i.e., during which the BNES stroke exceeds the half-width of the clearance band (approximately 30 mm) and the strongly nonlinear restoring force acts. The cords engage at all intensities considered, and increasingly more often as the intensity increases (from 8–26% of the time at 1.5 m/s to 12–40% at 3.5 m/s), so that the BNESs are activated throughout this range and their activation threshold lies below 1.5 m/s. Consistently, the reductions of the cumulative VIV-induced and drag energies remain positive at all intensities and increase with the intensity, from approximately 8–10% and 11–14% at 1.5–2.5 m/s to approximately 19% and 21–24% at 3–3.5 m/s, without any sign of degradation within this range.

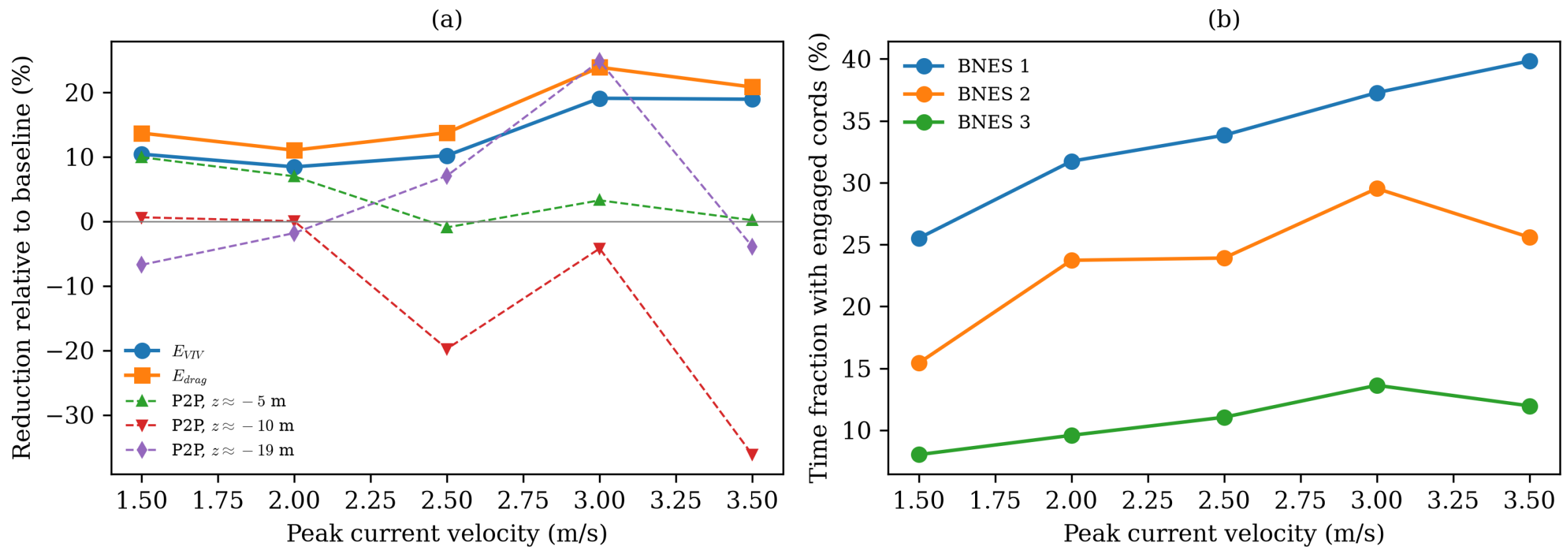

Figure 14. Working range of the BNESs optimized for current profile 2 (Table A.1) under the current of profile 2 scaled to peak velocities between 1.5 and 3.5 m/s: (a) Reductions of the cumulative VIV-induced and drag energies and of the P2P amplitudes at the three sensing locations relative to the baseline cable at the same intensity (negative values denote increases); (b) fraction of time during which the cords of each BNES are engaged (in tension), i.e., the strongly nonlinear restoring force is engaged.

The P2P amplitude reductions, in contrast, vary non-monotonically with the intensity: The downstream reduction is largest at the design intensity (25% at 3 m/s), whereas at the intermediate sensing location, located between two BNESs, the amplitude increases at 2.5 and 3.5 m/s. This variability of the point-wise amplitudes is consistent with their sensitivity to small perturbations discussed below (Fig. 15). Hence, while the energy benefits of the BNESs are robust and grow with the intensity of the current, as expected for energy-dependent nonlinear attachments (Manevitch et al., 2014; Romeo et al., 2015), the local amplitude benefits depend on the specific design and current, which suggests optimizing the BNESs over the entire expected range of currents.

The sensitivity of the dynamics, discussed in Section 4.3 in relation to chaotic cross-well oscillations, was assessed by repeating the simulations of the optimized designs and of the baseline cable for all three current profiles with the current amplitude perturbed by one part in a million. As shown in Fig. 15, the separation between the nominal and the perturbed cross-flow responses at $z \approx -19$ m remains below 1.5 mm for the baseline cable, whereas for the BNES-equipped cable it grows rapidly, to 6–60 mm within the 10 s simulation window, i.e., to the order of the vibration amplitude itself. This extreme sensitivity of the BNES-equipped system to infinitesimal perturbations is a characteristic signature of chaotic dynamics, and confirms that the cross-well transitions of the BNESs are, at least in part, chaotic. Despite the large point-wise differences, the cumulative VIV-induced and drag energies change by less than 4.2%, whereas the downstream P2P amplitude changes by up to 12%, which supports the use of the integral energy metrics (6) as robust measures of performance; on the other hand, point-wise amplitude metrics of BNES-equipped cables should be interpreted statistically, e.g., over multiple current realizations.

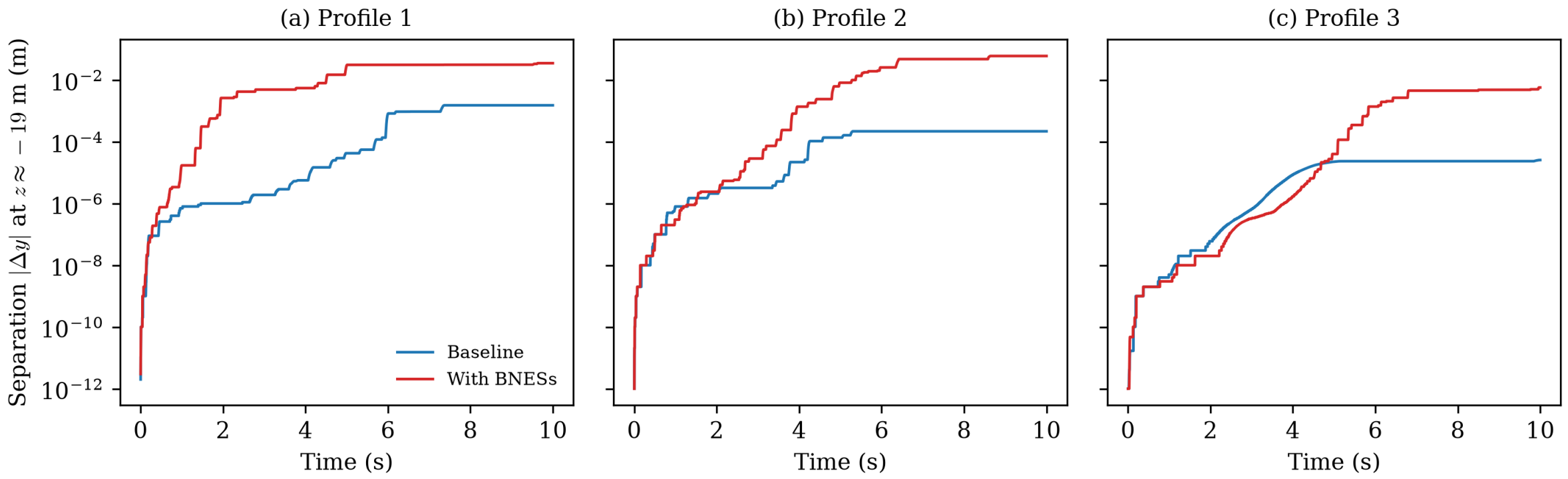

Figure 15. Sensitivity to a relative perturbation of $10^{-6}$ of the current amplitude for current profiles 1–3: Running maximum of the separation between the nominal and perturbed cross-flow displacements of the cable at $z \approx -19$ m, for the baseline cable and the cable with the optimized BNESs.

Finally, to isolate the contribution of bi-stability under the present VIV conditions, each optimized design (Tables 1, 2 and A.1) was compared with an equal-mass monostable NES with identical cord stiffness and damping, obtained by setting $\phi_0 = 0$, i.e., by setting the natural length of the cords equal to the half-length of the attachment rod, which eliminates bi-stability and the clearance band and yields a purely cubic stiffness law (Mojahed et al., 2018; Michaloliakos et al., 2025). The results are listed in Table 4. For all three current profiles the bi-stable design achieves larger reductions of the cumulative VIV-induced and drag energies than its monostable counterpart, by factors of approximately 1.3–2.2. The P2P reductions are mixed at individual sensing locations: The bi-stable design is clearly superior downstream for current profiles 1 and 2 and at the intermediate location for current profile 3, whereas the monostable NES performs slightly better at the two upstream locations for current profile 1, at the intermediate location for current profile 2 and downstream for current profile 3. These results show that the bi-stability specifically enhances the mitigation of the VIV energy intake and drag. It is noted that the monostable NESs were not separately optimized; an optimally tuned monostable NES and a tuned mass damper are left for future work.

Table 4. Reductions (%) relative to the baseline cable achieved by the optimized BNESs and by equal-mass monostable (cubic) NESs with the same cord properties ($\phi_0 = 0$).

| Profile | Device | $E_{VIV}$ | $E_{drag}$ | P2P, −5 m | P2P, −10 m | P2P, −19 m |
|---|---|---|---|---|---|---|
| 1 | BNES | 12.9 | 14.8 | 4.1 | 1.3 | 14.6 |
| 1 | Monostable | 9.1 | 11.2 | 5.9 | 2.7 | 1.8 |
| 2 | BNES | 19.2 | 23.9 | 3.6 | −4.1 | 25.1 |
| 2 | Monostable | 14.0 | 14.3 | 3.2 | −2.6 | 15.9 |
| 3 | BNES | 20.3 | 22.4 | 6.7 | 7.0 | 31.0 |
| 3 | Monostable | 10.0 | 10.0 | 4.4 | 0.6 | 35.0 |

## 5. Concluding Remarks

A comprehensive investigation was carried out into passive VIV mitigation of subsea power cables using a set of optimized strongly nonlinear bi-stable attachments, referred to as bi-stable nonlinear energy sinks (BNESs). These devices were implemented and their mitigation performance was evaluated within the open-source MoorDyn framework. Leveraging the modular structure and extensibility of MoorDyn, a fully 3D computational framework was developed, with capacity for capturing realistic cable dynamics, hydrodynamic forcing via Morison-type models, and nonlinear passive vibration-mitigation mechanisms under spatially and temporally varying dynamic currents. The unsteady and transient nature of the VIV-inducing dynamic current excitations considered in this work is emphasized at this point, which poses distinct challenges in mitigating the resulting highly non-stationary vibrations of the subsea cable. In this demanding load scenario, the BNESs proved to be efficacious in robustly and effectively suppressing cable VIV, demonstrating good performance as passive broadband vibration absorbers, energy scatterers and response mitigators.

A key contribution of this work is the integration of BNES devices into a time-domain cable dynamics computational solver that naturally accommodates large deformations, nonlinear hydrodynamics, and complex excitation scenarios, such as vortex-induced non-stationary vibrations. The developed modeling approach allows for the BNESs to interact with the cable in a physically consistent manner, preserving the essential bi-stable dynamics and enabling strong nonlinear energy exchanges and transfers between the flow and the cable, as well as within the cable itself under VIV excitation. Specifically, the introduction of rod-local slider constraints further ensured consistency between the analytical single-DOF B-NES models and the fully 3D numerical implementation of the integrated cable-BNESs system.

An extensive data-driven optimization study was conducted by systematically sampling the BNES design space across multiple dynamic current profiles. For each case, thousands of parallel simulation and processing iterations were performed, allowing for the exploration of trade-offs between competing performance objectives. The results consistently demonstrated that properly tuned BNES configurations can achieve substantial reductions in peak-to-peak cable vibration amplitudes at multiple locations along the cable. But perhaps equally important, the BNES mitigation performance was not limited to local amplitude reduction but also extended to significant decreases in cumulative VIV-induced and drag-induced energies of the cable, providing a new, more global and physically meaningful measure of VIV suppression. Hence, this work showed that the addition of optimized local strongly nonlinear vibration attachments to a subsea cable undergoing highly unsteady VIV, can significantly reduce the overall energy intake from the flow to the cable, but, perhaps even more noteworthy, decrease the drag amplification experienced in the cable due to VIV. To the authors' best knowledge, the capacity to reduce VIV-induced drag amplification in a subsea cable using passive nonlinear attachments is reported for the first time in the literature and can be particularly beneficial for reducing short- and long-term fatigue in dynamic cables.

Across the three examined dynamic current profiles, including low-intensity, high-intensity, and surface-dominated flow currents, the optimized BNESs exhibited robust performance. The mitigation effect was found to be particularly pronounced for the high-energy and strongly localized excitation scenario, highlighting a fundamental advantage of nonlinear passive devices over their linear counterparts (such as the tuned mass damper); namely, that the effectiveness of a BNES is enhanced with increasing energy, within the range of the current intensities considered herein, due to its inherently energy-dependent resonance dynamics. This feature makes the BNES especially attractive for mitigating extreme or highly non-stationary, multi-frequency transient VIV events that are most critical from fatigue and reliability standpoints.

Time–frequency analysis further clarified the physical nonlinear mechanisms underlying the reported effective VIV mitigation due to the actions of the optimized BNESs. Wavelet transforms of cable displacements, relative BNESs motions, and instantaneous VIV power input signals revealed a consistent pattern of targeted nonlinear energy transfers and VIV energy scattering from high-amplitude, low-frequency VIV modes (which are dominant in the baseline cable undergoing VIV), to lower-amplitude, higher frequency cable modes. This energy scattering (modal redistribution) facilitates rapid dissipation through hydrodynamic damping and internal structural losses of the cable vibrations, explaining the simultaneous reduction in vibration amplitudes and the cumulative energy intake in the cable from the flow. The realization of cross-well oscillations in the BNES responses confirmed the active role of bi-stability in sustaining broadband low-to-high frequency energy transfers.

A modal-band analysis, based on the natural frequencies of the cable and integrated over its entire span, confirmed that this redistribution corresponds to a transfer of VIV energy from the dominant low-order cable modes to higher-order modes. For current intensities between 1.5 and 3.5 m/s the BNESs remained activated (cords engaged) and the energy reductions increased with the current intensity, and the bi-stable design outperformed an equal-mass monostable (cubic) NES in terms of the cumulative energy metrics for all current profiles.

In synopsis, this study demonstrates that BNES devices, when properly designed and optimized, offer an effective, fully passive, and robust solution for mitigating VIV of subsea power cables under realistic unsteady operating conditions. The numerical and optimization framework developed herein provides a versatile platform for future investigations, including the study of combined current–wave loading, platform-induced motions; the optimization of BNES designs over a target range of current intensities and their cross-evaluation over the currents experienced during the service life of the cable; a rigorous characterization of the chaotic cross-well dynamics; comparisons with optimally tuned monostable NESs and tuned mass dampers under unsteady VIV conditions; fatigue life estimation and betterment through VIV reduction in these systems; and extension to other types of marine structures.

**Acknowledgements**

This research was funded in part by the National Offshore Wind Research and Development Consortium (NOWRDC), Contract Ref. No. 187. This support is gratefully acknowledged. The authors also greatly appreciate being granted access to the Hardware-Accelerated Learning (HAL) cluster, which is a High-Performance Computing (HPC) facility at the University of Illinois Urbana-Champaign. The authors would like to thank the reviewers for their insightful comments which led to revisions that made the contents of this paper more clear and complete.

## Appendix: Optimization results for dynamic current profile 2

The normalized performance maps for the numerical optimization subject to current profile 2 are depicted in Fig. A.1. The selected Pareto-optimal BNES design is indicated by a red marker, and the optimized system parameters for the three BNESs are listed in Table A.1.

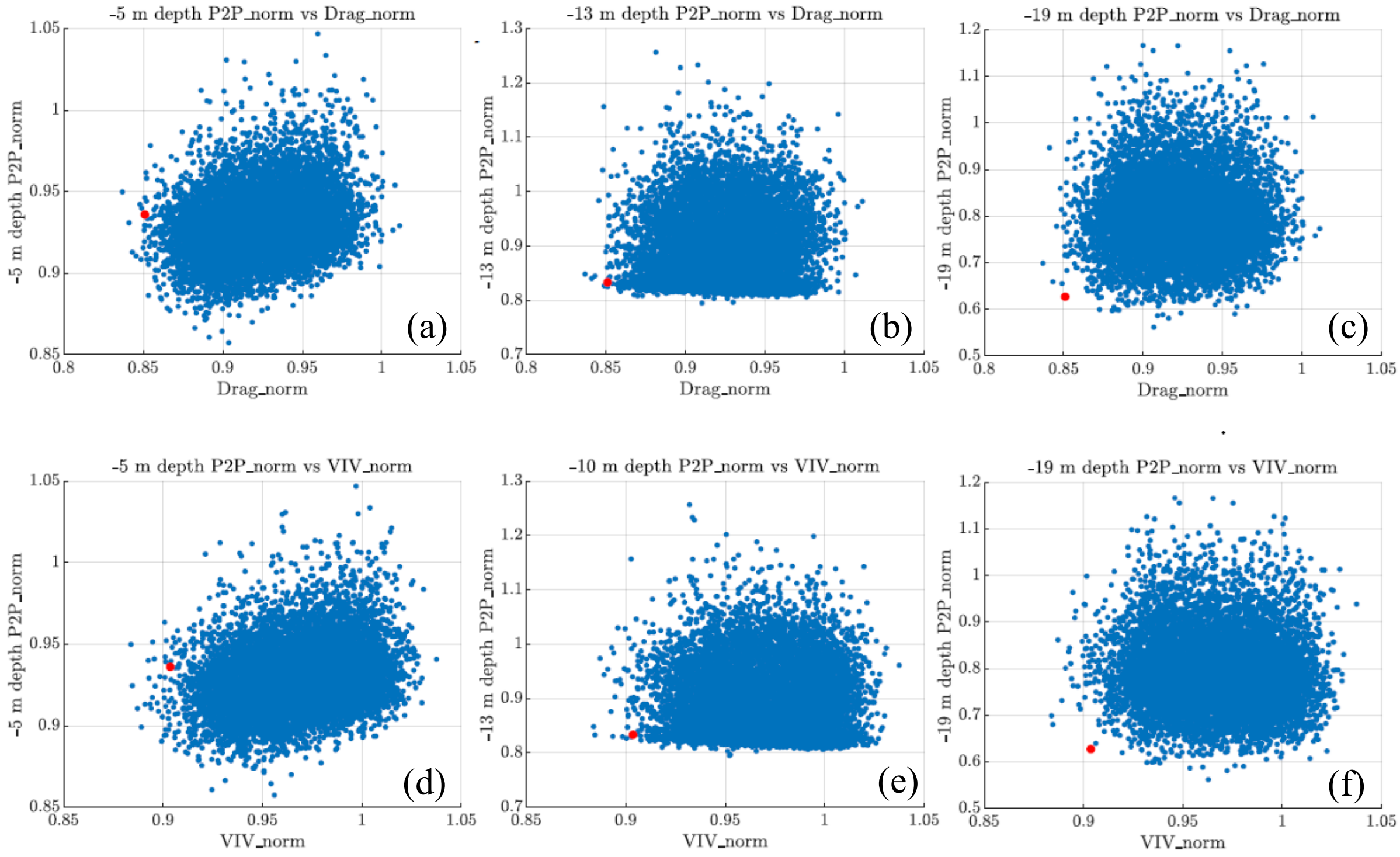

Figure A.1. Normalized performance maps for current profile 2: (a-c) P2P transverse VIV amplitude versus cumulative VIV-induced drag energy to the cable at the three sensing locations on the cable; (d–f) P2P transverse VIV amplitude versus cumulative VIV-induced energy intake to the cable (10,000 sampled BNES designs shown in each plot, and red marker denotes the selected Pareto-optimal BNES configuration).

Table A.1. Current profile 2: Optimal parameters of the BNESs (see Fig. 2a).

| BNES | $EA_i$ (N) | $d_i$ (Pa-s) | $\varphi_0$ (deg) | Mass (kg) |
|---|---|---|---|---|
| 1. Top | $1.13\times10^6$ | $3.00\times10^2$ | 3.562 | 4.94 |
| 2. Intermediate | $1.43\times10^6$ | $3.00\times10^2$ | 3.562 | 4.43 |
| 3. Bottom | 2.35×106 | $3.00\times10^2$ | 3.562 | 3.65 |